\documentclass[11pt,a4paper]{article}

\usepackage[utf8]{inputenc}
\usepackage[T1]{fontenc}
\usepackage{amsmath,amssymb}
\usepackage{booktabs}
\usepackage{tabularx}
\usepackage[margin=2.5cm]{geometry}

\usepackage{amsmath,amssymb,amsfonts}
\usepackage{amsthm}
\usepackage{mathrsfs}
\usepackage[table]{xcolor}

\usepackage{graphicx}
\usepackage{multirow}

\usepackage{textcomp}
\usepackage{manyfoot}
\usepackage[title]{appendix}
\usepackage{placeins}
\usepackage{listings}

\usepackage[numbers,sort&compress]{natbib}

\usepackage[
    colorlinks=true,
    linkcolor=magenta,  % ecuaciones, secciones y figuras
    citecolor=blue,   % citas bibliográficas [1,2,...]
    urlcolor=cyan!70!blue    % enlaces web
]{hyperref}

\theoremstyle{plain}

\theoremstyle{definition}

\theoremstyle{remark}

\usepackage{comment}

\title{\vspace{-1.5cm}
On the Metric Propagator and Affine Modes of Extended Hybrid Metric--Palatini Gravity with Ricci--Squared Invariants}

\author{
\small{
Jonathan Ramírez$^{\S}$\footnote{Corresponding author:
\href{mailto:ramirez.jsg@gmail.com}{ramirez.jsg@gmail.com}}
\qquad
Gustavo Melgarejo$^{\S}$\footnote{
\href{mailto:ga.melgarejoc@gmail.com}{ga.melgarejoc@gmail.com}}
}\\
\small{
\textit{Departamento de Física Teórica, Instituto de Física,}}\\
\textit{\small{Universidade do Estado do Rio de Janeiro,}}\\
\textit{\small{Rua São Francisco Xavier 524, Maracanã,}}\\
\textit{\small{CEP 20550-013, Rio de Janeiro, Brazil.}}\\[2pt]
\small{$^{\S}$These authors contributed equally to this work.}
}

\date{}

\usepackage{setspace}
\begin{document}

\maketitle

\begin{abstract}
We investigate an extended hybrid metric--Palatini theory defined by an
arbitrary function $f(R,\mathcal{R},\hat{Q},Q,\mathcal{Q})$, where $R$ and
$\mathcal{R}$ are the metric and Palatini scalar curvatures, and
$Q=R_{\mu\nu}R^{\mu\nu}$,
$\mathcal{Q}=\mathcal{R}_{(\mu\nu)}\mathcal{R}^{(\mu\nu)}$, and
$\hat{Q}=R_{\mu\nu}\mathcal{R}^{(\mu\nu)}$ are the metric, Palatini, and
mixed quadratic Ricci invariants, respectively. We derive the field equations and analyze the Minkowski weak--field limit. We obtain the metric propagator and identify the additional
spin--$2$ and spin--$0$ poles that couple to the metric source. When these
poles are finite, real, and simple, the spin-2 sector of the metric propagator contains
two additional poles, at least one of which has a negative
residue. Once the spin-2 sector is restricted to contain at most one additional pole, the absence of ghostlike metric-coupled excitations in the generic
two-scalar branch requires eliminating the remaining spin--$2$ pole,
whereas one healthy metric-coupled spin--$2$ pole may coexist with at most
one healthy scalar pole when the scalar denominator becomes linear in
$k^2$. We derive the corresponding algebraic conditions on the
Minkowski-background derivatives of $f$ and specialize them to hybrid
$f(R,\mathcal{R})$, mixed $f(R,\mathcal{Q})$, and purely Palatini
$f(\mathcal{R},\mathcal{Q})$ theories, including the metric $f(R)$ and
Palatini $f(\mathcal{R})$ limits. We further show that the complete linearized system can admit modes carried entirely by the independent connection that are absent from the metric propagator. These results therefore provide necessary flat-space
stability criteria for the metric-coupled sector on the branches analyzed,
supplemented by the analysis of purely affine modes.
\end{abstract}

\section{Introduction}

Extensions of General Relativity (GR) have been widely investigated in
connection with cosmological tensions, the dark sector, and the
inflationary epoch \cite{ABDALLA202249}. Among them, theories containing
higher-order curvature invariants provide a natural framework for
modifying gravitational dynamics
\cite{PhysRevD.71.063513,Stelle1978,Bajardi2024,Salvio2018,
NOJIRI2021115617}. Such modifications can introduce degrees of freedom
beyond the massless spin--$2$ graviton of GR. A necessary requirement
for their theoretical consistency is therefore that the additional
modes have acceptable masses and kinetic signs. Around Minkowski
spacetime, these properties can be investigated through the poles and
residues of the linearized propagator.

A particularly well-studied class of modified-gravity theories is
$f(R)$ gravity, in which the Einstein--Hilbert Lagrangian is replaced by a general function of the Ricci scalar \cite{Buchdahl1970}. Its dynamics depend
essentially on the choice of fundamental variables. In the metric
formalism, the metric $g_{\mu\nu}$ is the only independent gravitational
variable, whereas in the Palatini formalism the metric and an affine
connection are varied independently. For a Lagrangian linear in the
corresponding curvature scalar, both formulations reproduce GR, but
nonlinear choices generally lead to different dynamical and
phenomenological properties \cite{De_Felice_2010}. The metric and
Palatini constructions can be combined by allowing the action to depend
simultaneously on the metric curvature scalar $R$ and the Palatini
curvature scalar $\mathcal R$
\cite{PhysRevD.85.084016,PhysRevD.87.084031}. The resulting generalized
hybrid metric--Palatini theories generically admit two additional
scalar degrees of freedom nonminimally coupled to gravity. Their theoretical structure and phenomenology have been studied in several contexts \cite{Bombacigno2019}, including connections with other scalar--tensor theories \cite{Ramirez2026}, cosmological applications \cite{Rosa_2024,Rosa:2019ejh}, gravitational-wave phenomenology \cite{GomesRosaPinto2025}, and black-hole solutions \cite{Bronnikov2020HMPG}.

In this work, we consider a tensorial extension of generalized hybrid
metric--Palatini gravity in which the action depends not only on $R$
and $\mathcal R$, but also on the Ricci--squared invariants
\begin{equation*}
Q=R_{\mu\nu}R^{\mu\nu},
\qquad
\mathcal Q=\mathcal R_{(\mu\nu)}
\mathcal R^{(\mu\nu)},
\qquad
\hat Q=R_{\mu\nu}\mathcal R^{(\mu\nu)}.
\end{equation*}
These are, respectively, the metric, Palatini, and mixed quadratic
Ricci invariants. The dependence on the independent connection is
restricted to the symmetric part $\mathcal R_{(\mu\nu)}$ of the
Palatini Ricci tensor. This construction provides a natural tensorial
extension beyond the scalar-curvature hybrid sector and a suitable
setting in which to investigate additional spin--$2$ excitations. In metric higher-derivative gravity, Ricci--squared terms generically
produce a massive spin--$2$ pole with negative residue
\cite{Stelle1978,VANNIEUWENHUIZEN1973478}. In metric--affine theories, however, the independent connection may support additional modes. When matter couples only to the metric, some of these modes may not appear in the metric propagator.

Ricci-based metric--affine theories have previously been considered in
studies of nonsingular black holes~\cite{Olmo:2012}, cosmological
dynamics~\cite{Shahidi2025}, and the stability and particle content of
higher-curvature theories
\cite{BeltranJimenez2019,Jimenez2020,Annala_2023,BarkerMarzo2024}.
The latter analyses concern Ricci-type Palatini or metric--affine
theories and do not include the simultaneous dependence on the metric
and Palatini Ricci tensors considered here. A systematic analysis of
the linear metric response in the general hybrid theory
$f(R,\mathcal R,\hat Q,Q,\mathcal Q)$, supplemented by an examination
of its affine sector, has not yet been presented. We
address this problem by deriving the metric and connection field
equations and linearizing them around Minkowski spacetime. We then
obtain the propagator governing the linear response of the metric to
conserved matter sources and determine the masses and residues of its
additional spin--$2$ and spin--$0$ poles. In the following, we refer to
this object simply as the metric propagator.

For finite, real, and simple poles, the spin--$2$ part of the metric
propagator generically contains two additional poles, at least one of
which has negative residue. Reducing the tensor denominator to at most
one additional pole leads to two qualitatively different scalar
branches. On the generic two-scalar branch, the absence of ghostlike
metric-coupled poles requires the spin--$2$ correction to vanish. On
the reduced-order scalar branch, by contrast, a positive-residue
spin--$2$ pole may coexist with at most one healthy scalar pole. We
express the corresponding conditions in terms of derivatives of $f$
evaluated on the Minkowski background and specialize them to relevant
metric, Palatini, and hybrid subclasses.

We also analyze homogeneous solutions of the complete linearized
metric--connection system. Such solutions can be absent from the metric
propagator while remaining eigenmodes of the coupled equations. In
particular, the conditions that eliminate the additional
metric spin--$2$ poles can place the theory on a purely affine
spin--$2$ branch. The existence and mass of this mode, together with the condition for the absence of tachyonic instability, can be determined from the coupled field equations, but these equations alone do not determine whether the mode is ghostlike. Accordingly, the residue conditions obtained from this propagator should be understood as necessary flat-space stability conditions for the metric-coupled sector, supplemented by the separate affine-mode analysis, rather than as a complete proof of ghost freedom for the full metric--connection theory.

The paper is organized as follows. In Sec.~\ref{Action}, we introduce
the action and derive the metric and connection field equations. In
Sec.~\ref{Weak-field-limit}, we linearize the theory around Minkowski
spacetime and obtain the equations governing the metric response.
Section~\ref{Constraints} contains the metric propagator, the pole and
residue analysis, and the specialization to relevant subclasses.
Appendix~\ref{APPENDIX-A} discusses the pure-trace nature of torsion in
the symmetric-Ricci sector, while Appendix~\ref{APPENDIX-B} provides a
direct analysis of the subclasses for which
$f^{(0)}_{,\mathcal R}=0$. The affine modes of the complete
linearized system are examined separately in
Appendix~\ref{app:affine-modes}. Our conclusions are presented in
Sec.~\ref{conclusions}. We use $\kappa^2=8\pi G$, set $c=\hbar=1$, and
adopt the metric signature $(-1,+1,+1,+1)$.

\section{Action and field equations for \texorpdfstring{$f(R,\mathcal{R},\hat{Q},Q,\mathcal{Q})$}{f(R,Rcal,Qhat,Q,Qcal)} gravity}\label{Action}

The action for extended hybrid metric--Palatini gravity with Ricci--squared invariants is
\begin{equation}
    S=\frac{1}{2\kappa^2}\int d^4x\,\sqrt{-g}\,
    f\!\left(R,\mathcal{R},\hat{Q},Q,\mathcal{Q}\right)
    +S_m,
    \label{actionf}
\end{equation}
where $S_m=S_m[g_{\mu\nu},\psi]$ is the matter action, assumed to be independent of the affine connection and minimally coupled to $g_{\mu\nu}$. All indices are raised and lowered with $g_{\mu\nu}$. The metric Ricci scalar is $R=g^{\mu\nu}R_{\mu\nu}$, where the metric Ricci tensor is
\begin{equation}
    R_{\mu\nu}
    =\partial_{\lambda}\Gamma^{\lambda}{}_{\nu\mu}
     -\partial_{\nu}\Gamma^{\lambda}{}_{\lambda\mu}
     +\Gamma^{\sigma}{}_{\nu\mu}\Gamma^{\lambda}{}_{\lambda\sigma}
     -\Gamma^{\sigma}{}_{\lambda\mu}\Gamma^{\lambda}{}_{\nu\sigma},
\end{equation}
and $\Gamma^{\rho}{}_{\mu\nu}$ is the Levi--Civita connection of the metric. The Palatini Ricci scalar is
\begin{equation}
    \mathcal{R}
    =g^{\mu\nu}\mathcal{R}_{\mu\nu}
    =g^{\mu\nu}\mathcal{R}_{(\mu\nu)},
\end{equation}
where the Palatini Ricci tensor is constructed from an independent affine connection $\widetilde{\Gamma}^{\rho}{}_{\mu\nu}$ according to
\begin{equation}
    \mathcal{R}_{\mu\nu}
    =\partial_{\lambda}\widetilde{\Gamma}^{\lambda}{}_{\nu\mu}
     -\partial_{\nu}\widetilde{\Gamma}^{\lambda}{}_{\lambda\mu}
     +\widetilde{\Gamma}^{\sigma}{}_{\nu\mu}
      \widetilde{\Gamma}^{\lambda}{}_{\lambda\sigma}
     -\widetilde{\Gamma}^{\sigma}{}_{\lambda\mu}
      \widetilde{\Gamma}^{\lambda}{}_{\nu\sigma}.
\end{equation}

In addition to the scalar curvatures, we consider the quadratic invariants
\begin{equation}
    Q\equiv R_{\mu\nu}R^{\mu\nu},
    \qquad
    \mathcal{Q}\equiv
    \mathcal{R}_{(\mu\nu)}\mathcal{R}^{(\mu\nu)},
    \qquad
    \hat{Q}\equiv
    R_{\mu\nu}\mathcal{R}^{(\mu\nu)}.
\end{equation}
Since the action depends on the Palatini Ricci tensor only through $\mathcal{R}$, $\hat{Q}$, and $\mathcal{Q}$, the independent connection enters the action only through the symmetric part $\mathcal{R}_{(\mu\nu)}$.

The theory defined by Eq.~\eqref{actionf} extends the model introduced in Ref.~\cite{Koivisto2013} by including the quadratic Ricci invariants constructed separately from the metric and Palatini Ricci tensors. In particular, the purely metric term $R_{\mu\nu}R^{\mu\nu}$ and the purely Palatini term $\mathcal{R}_{(\mu\nu)}\mathcal{R}^{(\mu\nu)}$, which were absent from that formulation, are now explicitly incorporated. These invariants modify the dynamical structure of the theory and can introduce additional propagating modes. To the best of our knowledge, a hybrid metric--Palatini theory simultaneously encompassing all five invariants has not previously been analyzed in this general form.

Variation of the action \eqref{actionf} with respect to the metric yields
\begin{eqnarray}\label{field-equation-Q}
\nonumber
&& f_{,R}R_{\mu\nu}-\frac{1}{2}g_{\mu\nu}f
-\left(\nabla_{\mu}\nabla_{\nu}-g_{\mu\nu}\Box_g\right)f_{,R}
+f_{,\mathcal{R}}\mathcal{R}_{(\mu\nu)}
+f_{,\hat{Q}}R_{\mu}{}^\lambda\mathcal{R}_{(\nu\lambda)}
\\ \nonumber
&&+f_{,\hat{Q}}R_{\nu}{}^\lambda\mathcal{R}_{(\mu\lambda)}
+\frac{1}{2}\Box_g\!\left(f_{,\hat{Q}}\mathcal{R}_{(\mu\nu)}\right)
+\frac{1}{2}g_{\mu\nu}\nabla_\alpha\nabla_\beta
\!\left(f_{,\hat{Q}}\mathcal{R}^{(\alpha\beta)}\right)
-\frac{1}{2}\nabla^\lambda\nabla_{\mu}
\!\left(f_{,\hat{Q}}\mathcal{R}_{(\lambda\nu)}\right)
\\ \nonumber
&&-\frac{1}{2}\nabla^\lambda\nabla_{\nu}
\!\left(f_{,\hat{Q}}\mathcal{R}_{(\lambda\mu)}\right)
-2\,\nabla_{\lambda}\nabla_{(\mu}
\!\left(f_{,Q}R^{\lambda}{}_{\nu)}\right)
+g_{\mu\nu}\nabla_{\alpha}\nabla_{\beta}
\!\left(f_{,Q}R^{\alpha\beta}\right)
+\Box_g\!\left(f_{,Q}R_{\mu\nu}\right)
\\
&&+2f_{,Q}R_\mu{}^\lambda R_{\nu\lambda}
+2f_{,\mathcal{Q}}g^{\lambda\alpha}
\mathcal{R}_{(\mu\alpha)}\mathcal{R}_{(\nu\lambda)}
=\kappa^2T_{\mu\nu}.
\end{eqnarray}
Here, subscripts following a comma denote partial derivatives of $f$ with respect to its arguments, $\nabla_\mu$ is the covariant derivative associated with $g_{\mu\nu}$, and $\Box_g\equiv g^{\mu\nu}\nabla_\mu\nabla_\nu$. The matter stress--energy tensor is defined by
\begin{equation}
    T_{\mu\nu}
    \equiv
    -\frac{2}{\sqrt{-g}}
    \frac{\delta S_m}{\delta g^{\mu\nu}}.
\end{equation}

Variation of Eq.~\eqref{actionf} with respect to the independent connection gives
\begin{equation}
    \frac{1}{2\sqrt{-g}}
    \widetilde{\nabla}_{\alpha}
    \left(\sqrt{-g}\,q^{\mu\nu}\right)
    =
    q^{\mu\sigma}S^{\nu}{}_{\alpha\sigma}
    +q^{\mu\nu}S^{\sigma}{}_{\sigma\alpha}
    -\frac{1}{3}\delta^{\nu}_{\alpha}
    q^{\mu\rho}S^{\sigma}{}_{\sigma\rho},
    \label{eq:83}
\end{equation}
where
\begin{equation}
    q^{\mu\nu}\equiv
    f_{,\mathcal{R}}g^{\mu\nu}
    +f_{,\hat{Q}}R^{\mu\nu}
    +2f_{,\mathcal{Q}}\mathcal{R}^{(\mu\nu)}.
    \label{eq:q}
\end{equation}
Here, $\widetilde{\nabla}_{\alpha}$ denotes the covariant derivative associated with $\widetilde{\Gamma}^{\rho}{}_{\mu\nu}$, and $S^\lambda{}_{\mu\nu}\equiv\widetilde{\Gamma}^{\lambda}{}_{[\mu\nu]}$ is the torsion tensor.

Although no torsionless condition is imposed \emph{a priori} on the independent connection, the dependence of the gravitational action solely on $\mathcal{R}_{(\mu\nu)}$, together with the absence of matter couplings to the connection, renders the full action projectively invariant. As shown in Appendix~\ref{APPENDIX-A}, the connection field equations restrict the torsion to its trace, $S_\mu\equiv S^\lambda{}_{\lambda\mu}$, which is a projective gauge mode and may be set to zero without affecting the dynamics \cite{Alfonso_2017}. With this gauge choice, the connection becomes torsionless and its field equation reduces to
\begin{equation}
    \widetilde{\nabla}_{\alpha}
    \left(
    \sqrt{-g}\,q^{\mu\nu}
    \right)
    =0.
    \label{eq:84}
\end{equation}

On this nondegenerate branch, we introduce an auxiliary metric
$\tilde{g}_{\mu\nu}$ through
\begin{equation}
    \sqrt{-\tilde{g}}\,\tilde{g}^{\mu\nu}
    =\sqrt{-g}\,q^{\mu\nu}.
    \label{Gamma-ind}
\end{equation}
Equation~\eqref{eq:84} is then equivalent to
$\widetilde{\nabla}_{\alpha}
(\sqrt{-\tilde{g}}\,\tilde{g}^{\mu\nu})=0$.
Since the connection is torsionless, this compatibility condition implies that $\widetilde{\Gamma}^{\mu}{}_{\nu\lambda}$ is the Levi--Civita connection of $\tilde{g}_{\mu\nu}$:
\begin{equation}
    \widetilde{\Gamma}^{\mu}{}_{\nu\lambda}
    =\frac{1}{2}\tilde{g}^{\mu\rho}
    \left(
    \partial_\nu\tilde{g}_{\rho\lambda}
    +\partial_\lambda\tilde{g}_{\rho\nu}
    -\partial_\rho\tilde{g}_{\nu\lambda}
    \right).
    \label{gamma-palatini}
\end{equation}

In the general theory, however, $q^{\mu\nu}$, and hence $\tilde{g}_{\mu\nu}$, depends implicitly on the independent connection through $\mathcal{R}_{(\mu\nu)}$, both directly in Eq.~\eqref{eq:q} and through the curvature dependence of the derivatives of $f$. Equations~\eqref{Gamma-ind} and \eqref{gamma-palatini} therefore constitute an implicit connection equation rather than a closed solution for the connection in terms of the metric. In Sec.~\ref{Weak-field-limit}, we solve the connection equation perturbatively on the branch $f^{(0)}_{,\mathcal{R}}\neq0$, for which the background value of $q^{\mu\nu}$ is nondegenerate. The relevant $f^{(0)}_{,\mathcal R}=0$ subclasses considered below are treated directly on the $f^{(0)}_{,\mathcal Q}\neq0$ branch in
Appendix~\ref{APPENDIX-B}.

\section{Weak--field limit and masses of the metric-coupled modes}
\label{Weak-field-limit}
In this section, we derive the linearized field equations governing metric perturbations and identify the modes that appear in the metric response.\footnote{Modes associated with the independent connection are analyzed separately in Appendix~\ref{app:affine-modes}.} 

We consider small perturbations around Minkowski spacetime, so that the metric can be written as
\begin{equation}
    g_{\mu\nu}=\eta_{\mu\nu}+h_{\mu\nu},
    \label{lin-metric}
\end{equation}
where $\eta_{\mu\nu}$ is the Minkowski metric and $|h_{\mu\nu}|\ll1$. Up to first order in $h_{\mu\nu}$, we then have
\begin{equation}
    g^{\mu\nu}=\eta^{\mu\nu}-h^{\mu\nu},
    \label{lin-invermet}
\end{equation}
where indices are raised and lowered with $\eta_{\mu\nu}$, e.g., $h^{\mu\nu}=\eta^{\mu\sigma}\eta^{\nu\rho}h_{\sigma\rho}$. The Levi--Civita connection of the perturbed metric can be expanded as
\begin{equation}
    \Gamma^{\mu}{}_{\nu\lambda}
    =\frac{1}{2}\eta^{\mu\rho}
    \left(
    \partial_\nu h_{\rho\lambda}
    +\partial_\lambda h_{\rho\nu}
    -\partial_\rho h_{\nu\lambda}
    \right)
    +\mathcal{O}(h^2),
\end{equation}
so that, at this order, covariant derivatives associated with the metric reduce to partial derivatives. The linearized Ricci tensor reads
\begin{equation}
    R^{(1)}_{\mu\nu}
    =\partial_\sigma\partial_{(\mu}h^{\sigma}{}_{\nu)}
    -\frac{1}{2}\Box h_{\mu\nu}
    -\frac{1}{2}\partial_\mu\partial_\nu h,
\end{equation}
where $\Box\equiv\eta^{\mu\nu}\partial_\mu\partial_\nu$ and $h\equiv\eta^{\mu\nu}h_{\mu\nu}$.

The expansion of $f(R,\mathcal{R},\hat{Q},Q,\mathcal{Q})$ around the background is
\begin{equation}
    f=f^{(0)}
    +f_{,R}^{(0)}R^{(1)}
    +f_{,\mathcal{R}}^{(0)}\mathcal{R}^{(1)}
    +\mathcal{O}(h^2),
\end{equation}
and analogously for all its derivatives. Here $f^{(0)}$, $f^{(0)}_{,R}$, $f^{(0)}_{,\hat{Q}}$, $f^{(0)}_{,\mathcal{Q}}$, $f^{(0)}_{,Q}$, and $f^{(0)}_{,\mathcal{R}}$ denote constant background values evaluated on the zero-curvature Minkowski background, $R=\mathcal{R}=\hat{Q}=Q=\mathcal{Q}=0$, while $R^{(1)}$ and $\mathcal{R}^{(1)}$ denote the corresponding first-order curvature perturbations.

In what follows, the auxiliary-metric construction is used only when $f^{(0)}_{,\mathcal R}\neq0$. The subclasses with $f^{(0)}_{,\mathcal R}=0$ are instead analyzed directly in Appendix~\ref{APPENDIX-B}.

Let $q_{\mu\nu}$ denote the inverse matrix of $q^{\mu\nu}$, and define $q\equiv\det(q_{\mu\nu})$. Taking the determinant of Eq.~\eqref{Gamma-ind}, we find that $\sqrt{-\tilde{g}}=(\sqrt{-g})^2/\sqrt{-q}$. Therefore,
\begin{equation}
    \tilde{g}^{\mu\nu}
    =\frac{\sqrt{-q}}{\sqrt{-g}}\,q^{\mu\nu},
    \label{eq:conf-metric}
\end{equation}
with
\begin{equation}
    g=\det(g_{\mu\nu})=-1-h+\mathcal{O}(h^2).
    \label{det-g}
\end{equation}

Expanding \eqref{eq:q}, we obtain
\begin{eqnarray}
\nonumber
q^{\mu\nu}
&=&
f^{(0)}_{,\mathcal{R}}\eta^{\mu\nu}
-f^{(0)}_{,\mathcal{R}}h^{\mu\nu}
+f^{(0)}_{,R\mathcal{R}}R^{(1)}\eta^{\mu\nu}
+f^{(0)}_{,\mathcal{R}\mathcal{R}}
 \mathcal{R}^{(1)}\eta^{\mu\nu}
\\ &&
+f^{(0)}_{,\hat{Q}}R^{(1)\mu\nu}
+2f^{(0)}_{,\mathcal{Q}}
 \mathcal{R}^{(1)(\mu\nu)}
+\mathcal{O}(h^2).
\label{eq:expansion-q}
\end{eqnarray}

From this, we find for the determinant
\begin{equation}
    \frac{1}{q}
    =(f^{(0)}_{,\mathcal{R}})^4
    \Bigg[
    -1+h
    -\left(
    \frac{
    4f^{(0)}_{,R\mathcal{R}}
    +f^{(0)}_{,\hat{Q}}
    }{
    f^{(0)}_{,\mathcal{R}}
    }
    \right)R^{(1)}
    -2\left(
    \frac{
    2f^{(0)}_{,\mathcal{R}\mathcal{R}}
    +f^{(0)}_{,\mathcal{Q}}
    }{
    f^{(0)}_{,\mathcal{R}}
    }
    \right)\mathcal{R}^{(1)}
    \Bigg]
    +\mathcal{O}(h^2).
    \label{eq:1/det-q}
\end{equation}

Substituting Eqs.~\eqref{det-g}, \eqref{eq:expansion-q}, and \eqref{eq:1/det-q} into Eq.~\eqref{eq:conf-metric} gives the auxiliary metric in the form
\begin{equation}
    \tilde{g}^{\mu\nu}
    =(f^{(0)}_{,\mathcal{R}})^{-1}
    \left(
    \eta^{\mu\nu}-\tilde{h}^{\mu\nu}
    \right)
    +\mathcal{O}(h^2),
    \label{g-up}
\end{equation}
whose inverse reads
\begin{equation}
    \tilde{g}_{\mu\nu}
    =f^{(0)}_{,\mathcal{R}}
    \left(
    \eta_{\mu\nu}+\tilde{h}_{\mu\nu}
    \right)
    +\mathcal{O}(h^2),
    \label{g-down}
\end{equation}
where we have defined
\begin{equation}
    \tilde{h}^{\mu\nu}
    \equiv
    h^{\mu\nu}
    +(\mathcal{B}+\mathcal{D})
     \mathcal{R}^{(1)}\eta^{\mu\nu}
    -\mathcal{C}R^{(1)\mu\nu}
    -2\mathcal{D}\mathcal{R}^{(1)(\mu\nu)}
    +\frac{1}{2}(2\mathcal{A}+\mathcal{C})
     R^{(1)}\eta^{\mu\nu},
\end{equation}
and the following constants have been introduced for simplicity:
\begin{equation}
    \mathcal{A}
    =\frac{f^{(0)}_{,R\mathcal{R}}}
    {f^{(0)}_{,\mathcal{R}}},
    \qquad
    \mathcal{B}
    =\frac{f^{(0)}_{,\mathcal{R}\mathcal{R}}}
    {f^{(0)}_{,\mathcal{R}}},
    \qquad
    \mathcal{C}
    =\frac{f^{(0)}_{,\hat{Q}}}
    {f^{(0)}_{,\mathcal{R}}},
    \qquad
    \mathcal{D}
    =\frac{f^{(0)}_{,\mathcal{Q}}}
    {f^{(0)}_{,\mathcal{R}}},
    \qquad
    \mathcal{E}
    =\frac{f^{(0)}_{,Q}}
    {f^{(0)}_{,\mathcal{R}}}.
\end{equation}

Equations~\eqref{g-up} and \eqref{g-down} give the expanded independent connection
\begin{equation}
    \widetilde{\Gamma}^{\mu}{}_{\nu\lambda}
    =\frac{1}{2}\eta^{\mu\rho}
    \left(
    \partial_\nu\tilde{h}_{\rho\lambda}
    +\partial_\lambda\tilde{h}_{\rho\nu}
    -\partial_\rho\tilde{h}_{\nu\lambda}
    \right)
    +\mathcal{O}(h^2),
\end{equation}
from which the linearized Palatini Ricci tensor follows
\begin{eqnarray}
\nonumber
\mathcal{R}^{(1)}_{(\mu\nu)}
&=&
R^{(1)}_{\mu\nu}
-(\mathcal{B}+\mathcal{D})
 \partial_\mu\partial_\nu\mathcal{R}^{(1)}
-\tfrac12(2\mathcal{A}+\mathcal{C})
 \partial_\mu\partial_\nu R^{(1)}
+\tfrac{\mathcal{C}}{2}\Box R^{(1)}_{\mu\nu}
\\ &&
+\mathcal{D}\Box\mathcal{R}^{(1)}_{(\mu\nu)}
-\tfrac12(\mathcal{B}+\mathcal{D})
 \eta_{\mu\nu}\Box\mathcal{R}^{(1)}
-\tfrac14(2\mathcal{A}+\mathcal{C})
 \eta_{\mu\nu}\Box R^{(1)}.
\label{eq:Ricci-cal}
\end{eqnarray}

Therefore, the field equations \eqref{field-equation-Q}, expanded up to first order in the metric perturbation $h_{\mu\nu}$, take the form
\begin{eqnarray}
\nonumber
&&
f_{,R}^{(0)}R^{(1)}_{\mu\nu}
-\frac{1}{2}\eta_{\mu\nu}
\left(
f_{,R}^{(0)}R^{(1)}
+f_{,\mathcal{R}}^{(0)}\mathcal{R}^{(1)}
\right)
+f_{,\mathcal{R}}^{(0)}
 \mathcal{E}\Box R^{(1)}_{\mu\nu}
+\frac{1}{2}\eta_{\mu\nu}
\left(
2f^{(0)}_{,RR}
+f_{,\mathcal{R}}^{(0)}\mathcal{E}
\right)\Box R^{(1)}
\\
\nonumber
&&
-\frac{1}{2}f_{,\mathcal{R}}^{(0)}
 (2\mathcal{A}+\mathcal{C})
 \partial_\mu\partial_\nu\mathcal{R}^{(1)}
-\left(
f^{(0)}_{,RR}
+f_{,\mathcal{R}}^{(0)}\mathcal{E}
\right)
\partial_\mu\partial_\nu R^{(1)}
+\frac{1}{4}\eta_{\mu\nu}
 f_{,\mathcal{R}}^{(0)}
 (4\mathcal{A}+\mathcal{C})
 \Box\mathcal{R}^{(1)}
\\
&&
+\frac{\mathcal{C}}{2}f_{,\mathcal{R}}^{(0)}
 \Box\mathcal{R}^{(1)}_{(\mu\nu)}
+f_{,\mathcal{R}}^{(0)}
 \mathcal{R}^{(1)}_{(\mu\nu)}
-\frac{1}{2}f^{(0)}h_{\mu\nu}
=\kappa^2T^{(1)}_{\mu\nu}.
\label{Expand-field-eqQ2}
\end{eqnarray}
Here $T^{(1)}_{\mu\nu}$ denotes the linear perturbation of the stress--energy tensor.

Consistency of the expansion around Minkowski spacetime requires the background field equations to be satisfied. In vacuum, this gives
\begin{equation}
    f^{(0)}=0.
    \label{eq:minkowski-background-condition}
\end{equation}

Taking the traces of Eqs.~\eqref{eq:Ricci-cal} and \eqref{Expand-field-eqQ2}, and using Eq.~\eqref{eq:minkowski-background-condition}, we obtain
\begin{eqnarray}
\nonumber
&&
\mathcal{R}^{(1)}
=
\frac{1}{
f_{,\mathcal{R}}^{(0)}
\big(
3(\mathcal{A}+\mathcal{B})
+\mathcal{C}+2\mathcal{D}
\big)
}
\Bigg[
\Big(
(3\mathcal{A}+\mathcal{C})
f_{,\mathcal{R}}^{(0)}
-(3\mathcal{B}+2\mathcal{D})
f_{,R}^{(0)}
\Big)R^{(1)}
\\ &&
+\Big(
(3\mathcal{B}+2\mathcal{D})
(3f_{,RR}^{(0)}
 +2f_{,\mathcal{R}}^{(0)}\mathcal{E})
-f_{,\mathcal{R}}^{(0)}
 (3\mathcal{A}+\mathcal{C})^2
\Big)\Box R^{(1)}
-(3\mathcal{B}+2\mathcal{D})
 \kappa^2T^{(1)}
\Bigg].
\label{R-caligrafico}
\end{eqnarray}
Here we assume
\begin{equation}\label{surface-non-zero}
    3(\mathcal{A}+\mathcal{B})
    +\mathcal{C}+2\mathcal{D}\neq0.
\end{equation}
When the left-hand side of Eq.~\eqref{surface-non-zero} vanishes identically, Eq.~\eqref{R-caligrafico} cannot be used, and the corresponding branch may admit an affine scalar mode. This possibility is analyzed separately in Appendix~\ref{app:affine-modes}.

Equation~\eqref{R-caligrafico} expresses the Palatini curvature scalar $\mathcal{R}^{(1)}$ in terms of $R^{(1)}$, $\Box R^{(1)}$, and $T^{(1)}$. The linearized field equations, however, still contain the tensorial quantity $\mathcal{R}^{(1)}_{(\mu\nu)}$. To eliminate it without explicitly inverting the full relation \eqref{eq:Ricci-cal}, we rewrite Eq.~\eqref{eq:Ricci-cal} as
\begin{equation}
    (1-\mathcal{D}\Box)
    \mathcal{R}^{(1)}_{(\mu\nu)}
    =X_{\mu\nu},
    \label{eq-R-X}
\end{equation}
where
\begin{eqnarray}
X_{\mu\nu}
&=&
R^{(1)}_{\mu\nu}
+\frac{\mathcal{C}}{2}\Box R^{(1)}_{\mu\nu}
-\frac12(2\mathcal{A}+\mathcal{C})
 \partial_\mu\partial_\nu R^{(1)}
-(\mathcal{B}+\mathcal{D})
 \partial_\mu\partial_\nu\mathcal{R}^{(1)}
\\
\nonumber
&&
-\frac14(2\mathcal{A}+\mathcal{C})
 \eta_{\mu\nu}\Box R^{(1)}
-\frac12(\mathcal{B}+\mathcal{D})
 \eta_{\mu\nu}\Box\mathcal{R}^{(1)}.
\label{Xmunu}
\end{eqnarray}
The Palatini Ricci tensor appears in Eq.~\eqref{Expand-field-eqQ2} only through the combination
$\left(1+\frac{\mathcal{C}}{2}\Box\right)
\mathcal{R}^{(1)}_{(\mu\nu)}$.
It can therefore be eliminated by multiplying
Eq.~\eqref{Expand-field-eqQ2} by $(1-\mathcal{D}\Box)$ and using
Eq.~\eqref{eq-R-X}. Since this step involves multiplication by
$(1-\mathcal{D}\Box)$, it gives the equation governing the metric
response but does not determine all homogeneous solutions of the
complete metric--connection system. These solutions are analyzed
separately in Appendix~\ref{app:affine-modes}. Eliminating
$\mathcal{R}^{(1)}$ through Eq.~\eqref{R-caligrafico} and simplifying
the resulting equation with its trace, we obtain
\begin{equation}
    \frac{1}{b(\Box)}
    \Big\{
    2a(\Box)G^{(1)}_{\mu\nu}
    +\big[a(\Box)-c(\Box)-d(\Box)\big]
     \eta_{\mu\nu}R^{(1)}
    -\frac{e(\Box)-d(\Box)}{\Box}
     \partial_\mu\partial_\nu R^{(1)}
    \Big\}
    =
    2\kappa^2T^{(1)}_{\mu\nu},
    \label{eq:O-in-GR}
\end{equation}
where
\begin{equation}
    G^{(1)}_{\mu\nu}
    =R^{(1)}_{\mu\nu}
    -\frac{1}{2}\eta_{\mu\nu}R^{(1)}
\end{equation}
is the linearized Einstein tensor, and $a(\Box)$, $b(\Box)$, $c(\Box)$, $d(\Box)$, and $e(\Box)$ are given by
\begin{eqnarray}
\label{a-cuadrito}
a(\Box)
&=&
(f_{,R}^{(0)}+f_{,\mathcal{R}}^{(0)})
+\big(
f_{,\mathcal{R}}^{(0)}\mathcal{C}
-\mathcal{D}f_{,R}^{(0)}
+f_{,\mathcal{R}}^{(0)}\mathcal{E}
\big)\Box
+f_{,\mathcal{R}}^{(0)}
\left(
\tfrac{\mathcal{C}^2}{4}
-\mathcal{D}\mathcal{E}
\right)\Box^2,
\\
\label{b-cuadrito}
b(\Box)
&=&
1-\mathcal{D}\Box,
\\
\label{c-cuadrito}
c(\Box)
&=&
(f_{,R}^{(0)}+f_{,\mathcal{R}}^{(0)})
-2\Big[
2f_{,\mathcal{R}}^{(0)}\mathcal{A}
+f_{,RR}^{(0)}
-f_{,R}^{(0)}
 \left(
 \mathcal{B}+\tfrac{\mathcal{D}}{2}
 \right)
+\tfrac{f_{,\mathcal{R}}^{(0)}\mathcal{C}}{2}
+\tfrac{f_{,\mathcal{R}}^{(0)}\mathcal{E}}{2}
\Big]\Box
\\
\nonumber
&&
+2\Big[
\tfrac{f_{,\mathcal{R}}^{(0)}}{3}
 (3\mathcal{A}+\mathcal{C})^2
-3\mathcal{B}f_{,RR}^{(0)}
+\tfrac{f_{,\mathcal{R}}^{(0)}}{24}\mathcal{C}^2
-2f_{,\mathcal{R}}^{(0)}\mathcal{E}\mathcal{B}
-2\mathcal{D}f_{,RR}^{(0)}
-\tfrac{3}{2}f_{,\mathcal{R}}^{(0)}
 \mathcal{E}\mathcal{D}
\Big]\Box^2,
\\
d(\Box)
&=&
\frac{
(\mathcal{B}+\mathcal{D})
\big(a(\Box)-3c(\Box)\big)\Box
}{
1+(3\mathcal{B}+2\mathcal{D})\Box
},
\label{d-cuadrito}
\\
\nonumber
e(\Box)
&=&
2\Big[
f_{,\mathcal{R}}^{(0)}
 (2\mathcal{A}+\mathcal{C}+\mathcal{E})
+f_{,RR}^{(0)}
-f_{,R}^{(0)}
 (\mathcal{B}+\mathcal{D})
\Big]\Box
\\
&&
-2\Big[
\tfrac{1}{4}f_{,\mathcal{R}}^{(0)}
 (2\mathcal{A}+\mathcal{C})
 (6\mathcal{A}+\mathcal{C})
-(3\mathcal{B}+2\mathcal{D})
 f_{,RR}^{(0)}
-f_{,\mathcal{R}}^{(0)}
 (2\mathcal{B}+\mathcal{D})\mathcal{E}
\Big]\Box^2.
\end{eqnarray}

Since $\partial^\mu G^{(1)}_{\mu\nu}=0$ and
$c(\Box)+e(\Box)-a(\Box)=0$, the left-hand side of
Eq.~\eqref{eq:O-in-GR} is transverse. The linearized field equations
are therefore compatible with
$\partial^\mu T^{(1)}_{\mu\nu}=0$.

As a consistency check, consider the GR limit within the branch $f^{(0)}_{,\mathcal R}\neq0$. In this limit,
$3(\mathcal A+\mathcal B)+\mathcal C+2\mathcal D=0$.
Since Eq.~\eqref{R-caligrafico} was derived assuming that this combination is nonzero, it cannot be used to recover the GR limit. The limit must instead be taken directly at the level of the linearized equations. Setting
$f^{(0)}_{,\mathcal{R}\mathcal{R}}
=f^{(0)}_{,R\mathcal{R}}
=f^{(0)}_{,RR}
=f^{(0)}_{,Q}
=f^{(0)}_{,\hat{Q}}
=f^{(0)}_{,\mathcal{Q}}=0$
before eliminating $\mathcal R^{(1)}$ gives
$\mathcal R^{(1)}_{\mu\nu}=R^{(1)}_{\mu\nu}$.
Combining this result with the normalization
$f^{(0)}_{,R}+f^{(0)}_{,\mathcal R}=1$
yields the linearized Einstein equations, with
$a(\Box)=b(\Box)=c(\Box)=1$ and
$d(\Box)=e(\Box)=0$.

To identify the scalar modes explicitly, we take the trace of
Eq.~\eqref{eq:O-in-GR} and factorize the resulting operator. This yields
\begin{eqnarray}
\nonumber
&&
\frac{
(f_{,R}^{(0)}+f_{,\mathcal{R}}^{(0)})
(\Box-M^2_+)(\Box-M^2_-)R^{(1)}
}{
M^2_+M^2_-
\left[
1+(3\mathcal{B}+2\mathcal{D})\Box
\right]
}
=
-\kappa^2T^{(1)}.
\end{eqnarray}
The candidate scalar masses are
\begin{equation}
    M^2_{\pm}
    =
    \frac{
    2f_{,\mathcal{R}}^{(0)}
    (3\mathcal{A}+\mathcal{C}+\mathcal{E})
    -f_{,R}^{(0)}
    (3\mathcal{B}+2\mathcal{D})
    +3f_{,RR}^{(0)}
    \pm\mathcal{S}
    }{
    2\Big[
    f_{,\mathcal{R}}^{(0)}
    (3\mathcal{A}+\mathcal{C})^2
    -(3\mathcal{B}+2\mathcal{D})
    \left(
    2f_{,\mathcal{R}}^{(0)}\mathcal{E}
    +3f_{,RR}^{(0)}
    \right)
    \Big]
    },
    \label{masas-campos}
\end{equation}
with
\begin{eqnarray}
\nonumber
\mathcal{S}^2
&=&
-4(f_{,\mathcal{R}}^{(0)}+f_{,R}^{(0)})
f_{,\mathcal{R}}^{(0)}
(3\mathcal{A}+\mathcal{C})^2
\\
\nonumber
&&
+4(f_{,\mathcal{R}}^{(0)}+f_{,R}^{(0)})
(3\mathcal{B}+2\mathcal{D})
\left(
2f_{,\mathcal{R}}^{(0)}\mathcal{E}
+3f_{,RR}^{(0)}
\right)
\\
&&
+\Big[
(3\mathcal{B}+2\mathcal{D})f_{,R}^{(0)}
-2f_{,\mathcal{R}}^{(0)}
(3\mathcal{A}+\mathcal{C}+\mathcal{E})
-3f_{,RR}^{(0)}
\Big]^2.
\end{eqnarray}

The factorized expression assumes
$f_{,R}^{(0)}+f_{,\mathcal{R}}^{(0)}\neq0$
and a nonvanishing denominator in Eq.~\eqref{masas-campos}. On this
quadratic branch, $\mathcal{S}^2>0$ gives two distinct real candidate
scalar poles, whereas $\mathcal{S}^2=0$ gives a degenerate root and
$\mathcal{S}^2<0$ gives a complex-conjugate pair. If the denominator
vanishes, the scalar operator is reduced in order. The finite real
simple-pole sector and the relevant reduced-order case are analyzed in
Sec.~\ref{Constraints}.

\section{Constraints on the \texorpdfstring
{$f\left(R,\mathcal{R},\hat{Q},Q,\mathcal{Q}\right)$}
{f(R,Rcal,Qhat,Q,Qcal)} models from the metric propagator}
\label{Constraints}

In this section we derive the necessary flat-space conditions for the
additional poles appearing in the metric response to be free from
ghostlike and tachyonic instabilities.  Starting
from the linearized field equations \eqref{eq:O-in-GR}, we invert them
to obtain the metric propagator $\Pi^{\lambda\sigma}_{\mu\nu}$, defined
through
\begin{equation}
\Pi^{-1\,\lambda\sigma}_{\mu\nu}h_{\lambda\sigma}
=
2\kappa^2 T^{(1)}_{\mu\nu}.
\end{equation}
Following the formalism of
Refs.~\cite{Biswas2011,Koivisto2013,biswas2013}, we use the transverse
and longitudinal projectors in momentum space. For conserved sources, the
longitudinal contributions do not contribute, and the relevant part of
the propagator can be written as\footnote{We use
$\partial_\mu\rightarrow ik_{\mu}$, so that $\Box\rightarrow-k^2$; a
massive pole is located at $k^2=-m^2$.}
\begin{equation}
k^2\Pi
=
\frac{(1+k^2\mathcal{D})\,\mathcal{P}^2}{a(-k^2)}
+
\frac{\big[1-(3\mathcal{B}+2\mathcal{D})k^2\big]\,
\mathcal{P}^0_s}
{a(-k^2)-3c(-k^2)},
\label{propagator1}
\end{equation}
where $\mathcal{P}^2$ and $\mathcal{P}^0_s$ are, respectively, the
standard spin--2 and spin--0 projectors, while $a(\Box)$ and $c(\Box)$
are defined in Eqs.~\eqref{a-cuadrito} and~\eqref{c-cuadrito}.

It is useful to separate the GR contribution to the propagator.
Equation~\eqref{propagator1} can be rewritten as
\begin{equation}
\Pi
=
\Pi_{\mathrm{GR}}
+\Phi(-k^2)\,\mathcal{P}^2
+\Psi(-k^2)\,\mathcal{P}^0_s,
\label{propagator}
\end{equation}
where
\begin{equation}
\Pi_{\mathrm{GR}}
=
\frac{1}{k^2}
\left(
\mathcal{P}^2-\frac{1}{2}\mathcal{P}^0_s
\right)
\end{equation}
is the usual GR propagator in the conserved-source sector. The
functions $\Phi$ and $\Psi$ contain the corrections due to the extended
hybrid structure:
\begin{equation}
\Phi(-k^2)
=
\frac{(1+k^2\mathcal{D})-a(-k^2)}
{k^2a(-k^2)},
\end{equation}
and
\begin{equation}
\Psi(-k^2)
=
\frac{
2\big[1-(3\mathcal{B}+2\mathcal{D})k^2\big]
+a(-k^2)-3c(-k^2)}
{2k^2\big[a(-k^2)-3c(-k^2)\big]}.
\end{equation}

The derivation assumes
\begin{equation}
f^{(0)}_{,\mathcal R}\neq0,
\qquad
3(\mathcal A+\mathcal B)+\mathcal C+2\mathcal D\neq0,
\label{general-branch-assumptions}
\end{equation}
as required, respectively, by the definitions of
$\mathcal A,\mathcal B,\mathcal C,\mathcal D,\mathcal E$ and by the
elimination of $\mathcal R^{(1)}$ in Eq.~\eqref{R-caligrafico}. The $f^{(0)}_{,\mathcal R}=0$ subclasses considered below are treated in
Appendix~\ref{APPENDIX-B}, while the complementary surface defined by
the second condition is analyzed in
Appendix~\ref{app:affine-modes}. We also impose the normalization
\begin{equation}
f^{(0)}_{,\mathcal R}+f^{(0)}_{,R}=1,
\end{equation}
which recovers the standard Einstein propagator in the infrared limit.
Unless otherwise stated, the pole and residue conditions below apply
to finite, real, and simple poles.

The spin--2 correction can be written as
\begin{equation}
\Phi(-k^2)
=
\frac{\delta_{\Phi}-\lambda_{\Phi}k^{2}}
{1-\omega_{\Phi}k^{2}+\lambda_{\Phi}k^{4}},
\label{Phi-particulaQ}
\end{equation}
where
\hypertarget{53}{\begin{eqnarray}
\nonumber
\delta_{\Phi}
&=&
f^{(0)}_{,\mathcal{R}}
(\mathcal{C}+\mathcal{D}+\mathcal{E}),
\\
\lambda_{\Phi}
&=&
\frac{f^{(0)}_{,\mathcal{R}}}{4}
\left(\mathcal{C}^{2}-4\mathcal{D}\mathcal{E}\right),
\\
\nonumber
\omega_{\Phi}
&=&
f^{(0)}_{,\mathcal{R}}\mathcal{C}
-\mathcal{D}\big(1-f^{(0)}_{,\mathcal{R}}\big)
+f^{(0)}_{,\mathcal{R}}\mathcal{E}.
\end{eqnarray}}

When
\begin{equation}
\mathcal{C}^{2}-4\mathcal{D}\mathcal{E}\neq0,
\end{equation}
the denominator of $\Phi(-k^2)$ is quadratic in $k^2$. If its two roots
are real and distinct, their residues add up to $-1$. Therefore, if
both roots correspond to propagating spin--2 poles, at least one has
negative residue.

One denominator factor may instead cancel against the numerator of
Eq.~\eqref{Phi-particulaQ}. Such a cancellation requires
\begin{equation}
\delta_\Phi^2-\omega_\Phi\delta_\Phi+\lambda_\Phi=0.
\end{equation}
After removing the common factor, the remaining simple spin--2 pole
has residue $-1$ and is also ghostlike. Coincident roots give a
higher-order pole and are not covered by the simple-pole criteria.
Consequently, the absence of ghostlike spin--2 poles in the finite real
simple-pole sector requires
\begin{equation}
\mathcal{C}^{2}-4\mathcal{D}\mathcal{E}=0.
\label{condition-single-spin2}
\end{equation}

Under condition~\eqref{condition-single-spin2},
Eq.~\eqref{Phi-particulaQ} reduces to
\begin{equation}
\Phi(-k^2)
=
\frac{\delta_\Phi}{1-\omega_\Phi k^2}.
\end{equation}
If $\omega_\Phi\neq0$, there is one additional spin--2 pole with mass
and residue
\begin{equation}
m_2^2=-\frac{1}{\omega_\Phi},
\qquad
\rho_2=-\frac{\delta_\Phi}{\omega_\Phi}.
\label{spin2-mass-residue}
\end{equation}
This pole is non-tachyonic and has positive residue when $\omega_\Phi<0$ and $\delta_\Phi>0$.

For $\mathcal D\neq0$, condition~\eqref{condition-single-spin2} gives
$\mathcal E=\mathcal C^2/(4\mathcal D)$, and hence
\hypertarget{spin2-single-residue}{\begin{equation}
\rho_2
=
-\frac{
f^{(0)}_{,\mathcal R}(\mathcal C+2\mathcal D)^2}
{
f^{(0)}_{,\mathcal R}(\mathcal C+2\mathcal D)^2
-4\mathcal D^2
}.
\label{spin2-single-residue}
\end{equation}}
If $\mathcal D=0$, condition~\eqref{condition-single-spin2} implies
$\mathcal C=0$. In this case, whenever an additional spin--2 pole is
present, its residue is $\rho_2=-1$.

We now turn to the scalar sector. Under
condition~\eqref{condition-single-spin2}, the denominator of the scalar
correction takes the form
\begin{equation}
a(-k^2)-3c(-k^2)
=
-2+\sigma_1 k^2+\sigma_2 k^4,
\end{equation}
where
\hypertarget{sigmas}{\begin{eqnarray}
\label{sigma1}
\sigma_1
&=&
-12f^{(0)}_{,\mathcal R}\mathcal A
-6f^{(0)}_{,RR}
+6\big(1-f^{(0)}_{,\mathcal R}\big)\mathcal B
+4\mathcal D\big(1-f^{(0)}_{,\mathcal R}\big)
-4f^{(0)}_{,\mathcal R}\mathcal C
-4f^{(0)}_{,\mathcal R}\mathcal E,
\\[4pt]
\label{sigma2}
\sigma_2
&=&
-18f^{(0)}_{,\mathcal R}\mathcal A^2
-12f^{(0)}_{,\mathcal R}\mathcal A\mathcal C
+12f^{(0)}_{,\mathcal R}\mathcal B\mathcal E
+6(3\mathcal B+2\mathcal D)f^{(0)}_{,RR}.
\end{eqnarray}}
The scalar correction is therefore
\begin{equation}
\Psi(-k^2)
=
\frac{
\sigma_1-2(3\mathcal B+2\mathcal D)+\sigma_2 k^2}
{
2\big(-2+\sigma_1 k^2+\sigma_2 k^4\big)
}.
\label{Psi-sigma-form}
\end{equation}

For $\sigma_2\neq0$, the scalar denominator is quadratic in $k^2$,
with discriminant
\begin{equation}
\Delta_s
\equiv
\sigma_1^2+8\sigma_2.
\label{scalar-discriminant}
\end{equation}
Two real and distinct scalar poles require $\Delta_s>0$. Denoting their
residues by $\rho_\pm$, their product is
\begin{equation}
\rho_+\rho_-
=
-\frac{
f^{(0)}_{,\mathcal R}
\big[3(\mathcal A+\mathcal B)+\mathcal C+2\mathcal D\big]^2}
{\Delta_s}.
\label{scalar-residue-product}
\end{equation}
The squared factor in the numerator is nonzero on the branch defined
by Eq.~\eqref{general-branch-assumptions}. Therefore, if
$f^{(0)}_{,\mathcal R}>0$, the residues have opposite signs and one
scalar pole is ghostlike.

If $f^{(0)}_{,\mathcal R}<0$, the scalar residues have the same sign.
However, if an additional spin--2 pole is present,
Eq.~\eqref{spin2-single-residue} gives $\rho_2<0$ for
$\mathcal D\neq0$. For $\mathcal D=0$, any additional spin--2 pole has
$\rho_2=-1$. Thus, a positive-residue spin--2 pole cannot coexist with
two positive-residue scalar poles on this generic two-scalar branch.

If $\omega_\Phi=0$, the spin--2 correction contains no additional
pole. For $\mathcal D\neq0$, conditions
\eqref{condition-single-spin2} and $\omega_\Phi=0$ give
\begin{equation}
f^{(0)}_{,\mathcal R}
=
\frac{4\mathcal D^2}{(\mathcal C+2\mathcal D)^2}>0,
\end{equation}
which is incompatible with two positive scalar residues. For
$\mathcal D=0$, condition~\eqref{condition-single-spin2} gives
$\mathcal C=0$, while $\omega_\Phi=0$ and
$f^{(0)}_{,\mathcal R}\neq0$ imply $\mathcal E=0$ and hence
$\delta_\Phi=0$.

Consequently, on the generic branch with two real and distinct scalar
poles, the absence of ghostlike metric-coupled poles requires
\begin{equation}
\mathcal{C}^{2}-4\mathcal{D}\mathcal{E}=0,
\qquad
\mathcal{C}+\mathcal{D}+\mathcal{E}=0.
\label{spin2-ghost-free-conditions}
\end{equation}
These relations imply
\begin{equation}
\mathcal{C}=-2\mathcal{E},
\qquad
\mathcal{D}=\mathcal{E},
\end{equation}
and therefore
\begin{equation}
\Phi(-k^2)=0.
\end{equation}

To state the remaining scalar conditions explicitly, we impose
Eq.~\eqref{spin2-ghost-free-conditions}. Equation~\eqref{Psi-sigma-form}
can then be written as
\begin{equation}
\Psi(-k^2)
=
\frac{\delta_\Psi+\lambda_\Psi k^2}
{2\big(1+\omega_\Psi k^2+\lambda_\Psi k^4\big)},
\label{Psi-unfactorized}
\end{equation}
where
\hypertarget{78}{\begin{eqnarray}
\nonumber
\lambda_\Psi
&=&
f^{(0)}_{,\mathcal{R}}(3\mathcal{A}-2\mathcal{E})^{2}
-
(3\mathcal{B}+2\mathcal{E})
\big(
2f^{(0)}_{,\mathcal{R}}\mathcal{E}
+3f^{(0)}_{,RR}
\big),
\\[4pt]
\delta_\Psi
&=&
f^{(0)}_{,\mathcal{R}}
\big(6\mathcal{A}+3\mathcal{B}\big)
+3f^{(0)}_{,RR},
\\[4pt]
\nonumber
\omega_\Psi
&=&
6f^{(0)}_{,\mathcal{R}}\mathcal{A}
-3\mathcal{B}
\big(1-f^{(0)}_{,\mathcal{R}}\big)
-2\mathcal{E}
+3f^{(0)}_{,RR}.
\end{eqnarray}}
On this branch,
\begin{equation}
\sigma_1=-2\omega_\Psi,
\qquad
\sigma_2=-2\lambda_\Psi.
\end{equation}

For $\lambda_\Psi\neq0$, Eq.~\eqref{Psi-unfactorized} factorizes as
\begin{equation}
\Psi(-k^2)
=
\frac{\delta_\Psi+\lambda_\Psi k^{2}}
{2\lambda_\Psi
(k^2+\mathscr{M}^2_{+})(k^2+\mathscr{M}^2_{-})},
\label{Psi-particulaQ}
\end{equation}
where
\begin{equation}
\mathscr{M}^2_{\pm}
=
\frac{
\omega_\Psi
\pm
\sqrt{\omega_\Psi^2-4\lambda_\Psi}}
{2\lambda_\Psi}.
\end{equation}
The corresponding residues are
\begin{equation}
z_{\pm}
=
\frac{
\sqrt{\omega_\Psi^2-4\lambda_\Psi}
\pm(2\delta_\Psi-\omega_\Psi)}
{
4\sqrt{\omega_\Psi^2-4\lambda_\Psi}
}.
\end{equation}
Both scalar poles are non-tachyonic and have positive residues when
\begin{equation}
\mathscr{M}_{\pm}^{2}>0,
\qquad
z_{\pm}>0.
\end{equation}
For two real and distinct poles, these inequalities are equivalent to
\begin{equation}
\lambda_\Psi>0,
\qquad
\omega_\Psi>2\sqrt{\lambda_\Psi},
\qquad
\delta_\Psi^2-\omega_\Psi\delta_\Psi+\lambda_\Psi<0.
\label{generic-scalar-stability}
\end{equation}
The last combination simplifies to
\begin{equation}
\delta_\Psi^2-\omega_\Psi\delta_\Psi+\lambda_\Psi
=
9f^{(0)}_{,\mathcal R}
(\mathcal A+\mathcal B)^2.
\label{scalar-residue-identity}
\end{equation}
After imposing Eq.~\eqref{spin2-ghost-free-conditions}, the second
condition in Eq.~\eqref{general-branch-assumptions} becomes
$\mathcal A+\mathcal B\neq0$. Therefore, on the branch considered
here, the last inequality in Eq.~\eqref{generic-scalar-stability} is
equivalent to
\begin{equation}
f^{(0)}_{,\mathcal R}<0.
\end{equation}

We now consider the reduced-order scalar branch
\begin{equation}
\sigma_2=0.
\label{reduced-scalar-branch}
\end{equation}
In this case Eq.~\eqref{Psi-sigma-form} becomes
\begin{equation}
\Psi(-k^2)
=
\frac{\sigma_1-2(3\mathcal B+2\mathcal D)}
{2\big(-2+\sigma_1 k^2\big)}.
\end{equation}
If $\sigma_1\neq0$, the scalar sector contains one additional pole,
with
\hypertarget{reduced-scalar-mass-residue}{\begin{equation}
m_0^2=-\frac{2}{\sigma_1},
\qquad
\rho_0=
\frac{\sigma_1-2(3\mathcal B+2\mathcal D)}
{2\sigma_1}.
\label{reduced-scalar-mass-residue}
\end{equation}}
This pole is non-tachyonic and has positive residue when
\begin{equation}
\sigma_1<0,
\qquad
\rho_0>0.
\label{reduced-scalar-healthy}
\end{equation}
If an additional spin--2 pole is present, it must independently satisfy
\begin{equation}
\omega_\Phi<0,
\qquad
\rho_2=-\frac{\delta_\Phi}{\omega_\Phi}>0.
\label{reduced-spin2-healthy}
\end{equation}

\begin{table*}[t]
\centering
\caption{Metric-coupled poles and necessary stability conditions in
$f(R,\mathcal R,\hat Q,Q,\mathcal Q)$ gravity.}
\label{tab:pole-summary}
\small
\rowcolors{2}{gray!10}{white}
\renewcommand{\arraystretch}{1.3}
\begin{tabularx}{\textwidth}{@{}p{0.25\textwidth}
p{0.20\textwidth}X@{}}
\toprule
Branch & Additional metric-coupled poles & Stability result
\\
\midrule

$\mathcal C^2-4\mathcal D\mathcal E\neq0$
&
Two spin--2 poles and, generically, two spin--0 poles
&
At least one spin--2 residue is negative.
\\[3pt]

$\mathcal C^2-4\mathcal D\mathcal E=0$,
$\sigma_2\neq0$
&
At most one spin--2 pole and two spin--0 poles
&
Absence of ghostlike poles requires
$\mathcal C+\mathcal D+\mathcal E=0$, together with
$\hyperlink{78}{\lambda_\Psi>0}$,
$\hyperlink{78}{\omega_\Psi>2\sqrt{\lambda_\Psi}}$, and
$f_{,\mathcal R}^{(0)}<0$. In this case the spin--2 mode is absent.
\\[3pt]

$\mathcal C^2-4\mathcal D\mathcal E=0$,
$\sigma_2=0$
&
At most one spin--0 pole and at most one spin--2 pole
&
The scalar pole requires $\hyperlink{sigmas}{\sigma_1<0}$ and $\hyperlink{reduced-scalar-mass-residue}{\rho_0>0}$.
A healthy spin--2 pole may coexist if
$\hyperlink{53}{\omega_\Phi<0}$ and $\hyperlink{spin2-single-residue}{\rho_2>0}$.
\\

\bottomrule
\end{tabularx}
\end{table*}

Thus, unlike the generic two-scalar branch, the reduced-order condition
does not require $\Phi(-k^2)=0$. If $\omega_\Phi=0$, there is no
additional spin--2 pole. If $\sigma_1=0$, then $\Psi(-k^2)$ is constant and contains no additional scalar pole. If the numerator in Eq.~\eqref{reduced-scalar-mass-residue} vanishes, then $\Psi(-k^2)=0$ and the scalar pole is absent. On the subbranch $\Phi(-k^2)=0$, the relation $\sigma_2=-2\lambda_\Psi$ shows that
$\sigma_2=0$ is equivalent to $\lambda_\Psi=0$. The metric-coupled branches and their necessary stability conditions
are summarized in Table~\ref{tab:pole-summary}.

In the following subsections, the arguments displayed in $f$ denote
the curvature invariants retained in each subclass. For example,
$f(R,\hat Q,Q,\mathcal Q)$ is obtained by removing only the dependence
on $\mathcal R$.

\subsection{The \texorpdfstring
{$f(R,\hat{Q},Q,\mathcal{Q})$}{f(R,Qhat,Q,Qcal)} models}
\label{A}

In this subclass, the dependence on the Palatini curvature scalar
$\mathcal R$ is removed, so that
\begin{equation}
f^{(0)}_{,R\mathcal R}
=
f^{(0)}_{,\mathcal R\mathcal R}
=
f^{(0)}_{,\mathcal R}
=0.
\end{equation}
We normalize the massless graviton contribution by setting
$f^{(0)}_{,R}=1$.

Since $f^{(0)}_{,\mathcal R}=0$, this subclass lies outside the auxiliary-metric construction used in Sec.~\ref{Weak-field-limit}. We restrict the analysis to the case $f^{(0)}_{,\mathcal Q}\neq0$. The results below follow from the direct linearization of the metric and connection field equations presented in Appendix~\ref{APPENDIX-B}.

The metric propagator contains a ghostlike massive spin--2 pole unless
\begin{equation}
\left(f^{(0)}_{,\hat Q}\right)^2
-
4f^{(0)}_{,\mathcal Q}f^{(0)}_{,Q}
=0.
\label{condition-f(R,Q,Q,Q)}
\end{equation}
Under this condition, the spin--2 correction vanishes,
\begin{equation}
\Phi(-k^2)=0,
\end{equation}
and the scalar correction reduces to
\begin{equation}
\Psi(-k^2)
=
\frac{3f^{(0)}_{,RR}}
{2\left(1+3f^{(0)}_{,RR}k^2\right)}.
\end{equation}
For $f^{(0)}_{,RR}\neq0$, this expression contains one additional
scalar pole with mass
\begin{equation}
m^2
=
\left(3f^{(0)}_{,RR}\right)^{-1}.
\end{equation}
Its residue is positive, and the pole is non-tachyonic when
$f^{(0)}_{,RR}>0$. If $f^{(0)}_{,RR}=0$, the scalar correction
vanishes and no additional finite scalar pole is present.

Thus, after imposing
Eq.~\eqref{condition-f(R,Q,Q,Q)}, the scalar sector coincides with that
of metric $f(R)$ gravity at the linearized level. Moreover, on the
branch considered here, the connection equation fixes the Palatini
Ricci tensor algebraically in terms of the metric Ricci tensor, so no
additional finite purely affine symmetric-Ricci pole is supported; see
Appendix~\ref{app:affine-modes}. These results provide the necessary
flat-space stability conditions for the finite linearized poles but do
not establish nonlinear stability \cite{Belenchia2018}.

\subsection{The generalized hybrid
\texorpdfstring{$f(R,\mathcal{R})$}{f(R,Rcal)} models}

In this subclass, we set
\begin{equation}
f^{(0)}_{,Q}
=
f^{(0)}_{,\hat Q}
=
f^{(0)}_{,\mathcal Q}
=0
\end{equation}
and impose the normalization
\begin{equation}
f^{(0)}_{,\mathcal R}+f^{(0)}_{,R}=1.
\end{equation}
Therefore, $\mathcal C=\mathcal D=\mathcal E=0$, and the spin--2
correction vanishes,
\begin{equation}
\Phi(-k^2)=0.
\end{equation}
The additional metric-coupled poles consequently belong to the scalar
sector.

The assumptions of the general analysis,
Eq.~\eqref{general-branch-assumptions}, reduce to
\begin{equation}
f^{(0)}_{,\mathcal R}\neq0,
\qquad
\mathcal A+\mathcal B\neq0.
\label{hybrid-domain-condition}
\end{equation}
The second condition, $\mathcal A+\mathcal B\neq0$, is a restriction required by the procedure used to derive the general propagator and should not be interpreted as an additional stability condition.
The complementary case, $\mathcal A+\mathcal B=0$, is not covered by the general propagator derivation and is analyzed separately in Appendix~\ref{app:affine-modes}, where it may admit a purely affine scalar mode.

The scalar correction takes the form
\begin{equation}
\Psi(-k^2)
=
\frac{\alpha+\beta k^2}
{2\left(1+\gamma k^2+\beta k^4\right)},
\label{psi-hybr}
\end{equation}
where
\begin{eqnarray}
\nonumber
\alpha
&=&
3\left(
\mathcal B f^{(0)}_{,\mathcal R}
+f^{(0)}_{,RR}
+2\mathcal A f^{(0)}_{,\mathcal R}
\right),
\\[4pt]
\beta
&=&
9\left(
f^{(0)}_{,\mathcal R}\mathcal A^2
-\mathcal B f^{(0)}_{,RR}
\right),
\\[4pt]
\nonumber
\gamma
&=&
3\left(
2\mathcal A f^{(0)}_{,\mathcal R}
-\mathcal B\big(1-f^{(0)}_{,\mathcal R}\big)
+f^{(0)}_{,RR}
\right).
\end{eqnarray}

We first consider the generic case $\beta\neq0$. For two real and
distinct poles, Eq.~\eqref{psi-hybr} gives the scalar masses
\begin{equation}
m^2_{\pm}
=
\frac{\gamma\pm\sqrt{\gamma^2-4\beta}}
{2\beta},
\label{masas-hyb-2}
\end{equation}
with residues
\begin{equation}
r_\pm
=
\frac{
\sqrt{\gamma^2-4\beta}
\pm(2\alpha-\gamma)}
{4\sqrt{\gamma^2-4\beta}}.
\end{equation}
Both poles are non-tachyonic and have positive residues when
\begin{equation}
\beta>0,
\qquad
\gamma>2\sqrt{\beta},
\qquad
\alpha^2-\gamma\alpha+\beta<0.
\label{conditions-hybrid-abg}
\end{equation}

Using the definitions of $\mathcal A$ and $\mathcal B$, the
combinations entering Eq.~\eqref{conditions-hybrid-abg} can be written
as
\begin{equation}
\alpha^2-\gamma\alpha+\beta
=
9f^{(0)}_{,\mathcal R}
(\mathcal A+\mathcal B)^2
=
\frac{
9\left(
f^{(0)}_{,R\mathcal R}
+
f^{(0)}_{,\mathcal R\mathcal R}
\right)^2}
{f^{(0)}_{,\mathcal R}},
\label{hybrid-residue-combination}
\end{equation}
and
\begin{equation}
\beta
=
\frac{
9\left[
\left(f^{(0)}_{,R\mathcal R}\right)^2
-
f^{(0)}_{,\mathcal R\mathcal R}
f^{(0)}_{,RR}
\right]}
{f^{(0)}_{,\mathcal R}}.
\label{beta-hybrid-derivatives}
\end{equation}
Since Eq.~\eqref{hybrid-domain-condition} holds, the last inequality
in Eq.~\eqref{conditions-hybrid-abg} is equivalent to
$f^{(0)}_{,\mathcal R}<0$. The conditions for the two scalar poles to
be non-tachyonic and have positive residues can therefore be expressed
as
\begin{equation}
f^{(0)}_{,RR}>0,
\qquad
f^{(0)}_{,\mathcal R\mathcal R}>0,
\qquad
f^{(0)}_{,\mathcal R}<0,
\qquad
\left(f^{(0)}_{,R\mathcal R}\right)^2
<
f^{(0)}_{,\mathcal R\mathcal R}
f^{(0)}_{,RR},
\label{conditions-hybrid-derivatives}
\end{equation}
together with the domain condition
\eqref{hybrid-domain-condition}.

We now consider the reduced-order case
\begin{equation}
\beta=0,
\label{hybrid-reduced-order}
\end{equation}
which, from Eq.~\eqref{beta-hybrid-derivatives}, is equivalent to
\begin{equation}
\left(f^{(0)}_{,R\mathcal R}\right)^2
=
f^{(0)}_{,\mathcal R\mathcal R}
f^{(0)}_{,RR}.
\label{hybrid-reduced-derivatives}
\end{equation}
In this case, Eq.~\eqref{psi-hybr} reduces to
\begin{equation}
\Psi(-k^2)
=
\frac{\alpha}
{2\left(1+\gamma k^2\right)}.
\label{psi-hybrid-reduced}
\end{equation}
For $\gamma\neq0$, the scalar sector contains one additional pole with
mass and residue
\begin{equation}
m_0^2=\frac{1}{\gamma},
\qquad
r_0=\frac{\alpha}{2\gamma}.
\label{hybrid-reduced-mass-residue}
\end{equation}
This pole is non-tachyonic and has positive residue when
\begin{equation}
\gamma>0,
\qquad
\alpha>0.
\label{hybrid-reduced-stability}
\end{equation}

Within the domain \eqref{hybrid-domain-condition},
Eq.~\eqref{hybrid-reduced-derivatives} implies
$f^{(0)}_{,\mathcal R\mathcal R}\neq0$. The coefficients $\alpha$ and
$\gamma$ can therefore be written as
\begin{eqnarray}
\alpha
&=&
\frac{
3\left(
f^{(0)}_{,R\mathcal R}
+
f^{(0)}_{,\mathcal R\mathcal R}
\right)^2}
{f^{(0)}_{,\mathcal R\mathcal R}},
\\[4pt]
\gamma
&=&
3\left[
\frac{
\left(
f^{(0)}_{,R\mathcal R}
+
f^{(0)}_{,\mathcal R\mathcal R}
\right)^2}
{f^{(0)}_{,\mathcal R\mathcal R}}
-
\frac{
f^{(0)}_{,\mathcal R\mathcal R}}
{f^{(0)}_{,\mathcal R}}
\right].
\end{eqnarray}
The conditions \eqref{hybrid-reduced-stability} are consequently
equivalent to
\begin{equation}
f^{(0)}_{,\mathcal R\mathcal R}>0,
\qquad
\left(
1+
\frac{f^{(0)}_{,R\mathcal R}}
     {f^{(0)}_{,\mathcal R\mathcal R}}
\right)^2
>
\frac{1}{f^{(0)}_{,\mathcal R}}.
\label{hybrid-reduced-stability-derivatives}
\end{equation}
Together with $f^{(0)}_{,\mathcal R\mathcal R}>0$,
Eq.~\eqref{hybrid-reduced-derivatives} also implies
$f^{(0)}_{,RR}\geq0$.

If $\gamma=0$, Eq.~\eqref{psi-hybrid-reduced} is momentum independent
and contains no additional scalar pole. Thus, within the domain of the
general propagator, the reduced-order branch contains at most one
finite metric-coupled scalar pole.

The generic conditions
\eqref{conditions-hybrid-derivatives} are consistent with the
weak--field scalar analysis of generalized hybrid
$f(R,\mathcal R)$ gravity in Ref.~\cite{Bombacigno2019}.

\subsubsection{The hybrid
\texorpdfstring{$R+f(\mathcal{R})$}{R+f(Rcal)} models}

We now consider the restricted hybrid model
$f(R,\mathcal{R})=R+f(\mathcal{R})$. In this case,
\begin{equation}
f^{(0)}_{,RR}
=
f^{(0)}_{,R\mathcal R}
=
f^{(0)}_{,Q}
=
f^{(0)}_{,\hat Q}
=
f^{(0)}_{,\mathcal Q}
=0.
\end{equation}
Here the coefficient of $R$ is fixed to unity, so we leave the
normalization of the massless propagator explicit rather than imposing
$f^{(0)}_{,R}+f^{(0)}_{,\mathcal R}=1$. The regular-branch derivation
also assumes
\begin{equation}
f^{(0)}_{,\mathcal R}\neq0,
\qquad
f^{(0)}_{,\mathcal R\mathcal R}\neq0,
\end{equation}
so that $\mathcal B
=f^{(0)}_{,\mathcal R\mathcal R}/f^{(0)}_{,\mathcal R}$ is well
defined and nonzero.

The propagator functions in the spin--2 and spin--0 sectors are
\begin{equation}
\Phi(-k^2)
=
-\frac{f^{(0)}_{,\mathcal R}}
{k^2\big(1+f^{(0)}_{,\mathcal R}\big)},
\end{equation}
and
\begin{equation}
\Psi(-k^2)
=
\frac{f^{(0)}_{,\mathcal R}}
{2\big(1+f^{(0)}_{,\mathcal R}\big)}
\left[
\frac{1}{k^2}
-
\left(
k^2-\frac{1+f^{(0)}_{,\mathcal R}}{3\mathcal B}
\right)^{-1}
\right].
\end{equation}

Combining the massless terms in $\Phi$ and $\Psi$ with the corresponding
terms in $\Pi_{\mathrm{GR}}$, we find that the GR propagator is
multiplied by
$\big(1+f^{(0)}_{,\mathcal R}\big)^{-1}$. This introduces no
additional pole and only rescales the effective Newton constant.

The remaining term in $\Psi$ contains one additional scalar pole,
consistent with the scalar--tensor representation of these models
\cite{Capo2015}. Its mass and residue are
\begin{equation}
m_0^2
=
-\frac{
f^{(0)}_{,\mathcal R}
\big(1+f^{(0)}_{,\mathcal R}\big)}
{3f^{(0)}_{,\mathcal R\mathcal R}},
\end{equation}
and
\begin{equation}
r
=
-\frac{f^{(0)}_{,\mathcal R}}
{2\big(1+f^{(0)}_{,\mathcal R}\big)}.
\end{equation}

Requiring a positive effective Newton constant, a non-tachyonic scalar
pole, and a positive scalar residue gives
\begin{equation}
f^{(0)}_{,\mathcal R\mathcal R}>0,
\qquad
-1<f^{(0)}_{,\mathcal R}<0.
\end{equation}
Thus, within the regular branch considered here, the finite
metric-coupled poles are stable in this parameter range. If
$f^{(0)}_{,\mathcal R\mathcal R}=0$, the additional scalar pole is
absent and the linearized propagator reduces to the GR propagator with
a rescaled Newton constant.

\subsection{The
\texorpdfstring{$f(R,\mathcal{Q})$}{f(R,Qcal)} models}
\label{sec:fRQcal}

In this subclass, we set
\begin{equation}
f^{(0)}_{,\mathcal R}
=
f^{(0)}_{,\mathcal R\mathcal R}
=
f^{(0)}_{,R\mathcal R}
=
f^{(0)}_{,Q}
=
f^{(0)}_{,\hat Q}
=0,
\qquad
f^{(0)}_{,R}=1.
\end{equation}

Since $f^{(0)}_{,\mathcal R}=0$, this subclass lies outside the
auxiliary-metric construction used in
Sec.~\ref{Weak-field-limit}. We restrict the analysis to the case
$f^{(0)}_{,\mathcal Q}\neq0$. The direct linearization presented in
Appendix~\ref{APPENDIX-B} then gives
\begin{equation}
\mathcal R^{(1)}_{(\mu\nu)}=0,
\end{equation}
and the metric field equation reduces to that of linearized metric
$f(R)$ gravity.

Accordingly, the spin--2 correction vanishes,
\begin{equation}
\Phi(-k^2)=0,
\end{equation}
while the scalar correction is
\begin{equation}
\Psi(-k^2)
=
\frac{3f^{(0)}_{,RR}}
{2\left(1+3f^{(0)}_{,RR}k^2\right)}.
\end{equation}
For $f^{(0)}_{,RR}\neq0$, the metric propagator contains one additional
scalar pole with mass
\begin{equation}
m^2
=
\left(3f^{(0)}_{,RR}\right)^{-1}.
\end{equation}
Its residue is positive, and the pole is non-tachyonic when
\begin{equation}
f^{(0)}_{,RR}>0.
\end{equation}
If $f^{(0)}_{,RR}=0$, the scalar correction vanishes and no additional
finite scalar pole is present.

Thus, on the branch considered here, the Palatini Ricci--squared
invariant $\mathcal Q$ introduces no additional finite metric-coupled
or purely affine symmetric-Ricci pole. The linearized pole structure
therefore coincides with that of metric $f(R)$ gravity. These
flat-space conditions do not, by themselves, establish stability on
curved backgrounds or at the nonlinear level
\cite{Belenchia2018}.

\subsection{The quadratic Palatini gravity
\texorpdfstring{$f(\mathcal{R},\mathcal{Q})$}
{f(Rcal,Qcal)} models}

These models have previously been studied in
Ref.~\cite{Olmo:2012}. In this subclass, we set
\begin{equation}
f^{(0)}_{,R}
=
f^{(0)}_{,RR}
=
f^{(0)}_{,R\mathcal R}
=
f^{(0)}_{,Q}
=
f^{(0)}_{,\hat Q}
=0,
\qquad
f^{(0)}_{,\mathcal R}=1.
\end{equation}

The corrections to the metric propagator are
\begin{equation}
\Phi(-k^2)
=
f^{(0)}_{,\mathcal Q},
\qquad
\Psi(-k^2)
=
\frac{3}{2}f^{(0)}_{,\mathcal R\mathcal R}
+f^{(0)}_{,\mathcal Q}.
\end{equation}
Both corrections are constants and therefore contain no additional finite poles.

Consequently, the metric propagator contains no additional propagating
pole around Minkowski spacetime. The direct analysis of the coupled
metric--connection system also shows that this subclass supports no
finite purely affine curvature pole; see
Appendix~\ref{app:affine-modes}. Thus, at the linearized level, its
finite-curvature spectrum contains no additional modes beyond the
massless graviton of GR.

\subsection{The metric \texorpdfstring{$f(R)$}{f(R)} and Palatini
\texorpdfstring{$f(\mathcal{R})$}{f(Rcal)} models}

In the pure metric formulation, we set
\begin{equation}
f^{(0)}_{,\mathcal R}
=
f^{(0)}_{,\mathcal R\mathcal R}
=
f^{(0)}_{,R\mathcal R}
=
f^{(0)}_{,Q}
=
f^{(0)}_{,\hat Q}
=
f^{(0)}_{,\mathcal Q}
=0,
\qquad
f^{(0)}_{,R}=1.
\end{equation}

Since $f^{(0)}_{,\mathcal R}=0$, this case lies outside the
auxiliary-metric construction of Sec.~\ref{Weak-field-limit}. The
propagator is instead obtained directly from the metric field
equations and agrees with the standard metric $f(R)$ result
\cite{Sotiriou:2008rp,NOJIRI20171}.

The spin--2 correction vanishes,
\begin{equation}
\Phi(-k^2)=0,
\end{equation}
while the scalar correction is
\begin{equation}
\Psi(-k^2)
=
\frac{3f^{(0)}_{,RR}}
{2\left(1+3f^{(0)}_{,RR}k^2\right)}.
\end{equation}
For $f^{(0)}_{,RR}\neq0$, the metric propagator contains one additional
scalar pole with mass
\begin{equation}
m^2
=
\left(3f^{(0)}_{,RR}\right)^{-1}.
\end{equation}
Its residue is positive, and the pole is non-tachyonic when
\begin{equation}
f^{(0)}_{,RR}>0.
\end{equation}
If $f^{(0)}_{,RR}=0$, the scalar correction vanishes and the
linearized spectrum reduces to that of GR.

In the Palatini formulation, we instead set
\begin{equation}
f^{(0)}_{,R}
=
f^{(0)}_{,RR}
=
f^{(0)}_{,R\mathcal R}
=
f^{(0)}_{,Q}
=
f^{(0)}_{,\hat Q}
=
f^{(0)}_{,\mathcal Q}
=0,
\qquad
f^{(0)}_{,\mathcal R}=1.
\end{equation}
The propagator corrections are then
\begin{equation}
\Phi(-k^2)=0,
\qquad
\Psi(-k^2)
=
\frac{3}{2}f^{(0)}_{,\mathcal R\mathcal R}.
\end{equation}
The constant scalar correction has no pole.
Consequently, around Minkowski spacetime, Palatini $f(\mathcal R)$
gravity propagates only the massless graviton of GR.
Table~\ref{tab:viability} summarizes the results of this section.

\begin{table}[!htbp]
\centering
\caption{Additional poles in the metric propagator and necessary flat-space
stability conditions for the relevant models.\\}
\label{tab:viability}
\setlength{\tabcolsep}{3pt}
\renewcommand{\arraystretch}{1.5}
\small
\rowcolors{2}{gray!10}{white}
\begin{tabularx}{\textwidth}{@{}
>{\raggedright\arraybackslash}p{0.26\textwidth}
>{\raggedright\arraybackslash}p{0.27\textwidth}
X@{}}
\toprule
\textbf{Model or branch}
& \textbf{Additional finite poles}
& \textbf{Necessary stability conditions} \\
\midrule

$f(R)$
&
At most one scalar pole
&
If present, $f^{(0)}_{,RR}>0$
\\[4pt]

$f(\mathcal R)$
&
No additional poles
&
No pole condition
\\[4pt]

$f(R,\mathcal R)$, $\beta\neq0$
&
Two scalar poles
&
\begin{tabular}[t]{@{}l@{}}
$f^{(0)}_{,RR}>0,\quad
 f^{(0)}_{,\mathcal R\mathcal R}>0,\quad
 f^{(0)}_{,\mathcal R}<0,$\\~~~~
$\left(f^{(0)}_{,R\mathcal R}\right)^2
<
f^{(0)}_{,\mathcal R\mathcal R}f^{(0)}_{,RR}$
\end{tabular}
\\[7pt]

$f(R,\mathcal R)$, $\beta=0$
&
At most one scalar pole
&
\begin{tabular}[t]{@{}l@{}}
If present, $f^{(0)}_{,\mathcal R\mathcal R}>0$ and\\
$\displaystyle
\left(
1+\frac{f^{(0)}_{,R\mathcal R}}
{f^{(0)}_{,\mathcal R\mathcal R}}
\right)^2
>
\frac{1}{f^{(0)}_{,\mathcal R}}$
\end{tabular}
\\[7pt]

$R+f(\mathcal R)$
&
At most one scalar pole
&
\begin{tabular}[t]{@{}l@{}}
For $-1<f^{(0)}_{,\mathcal R}<0$ and
$f^{(0)}_{,\mathcal R\mathcal R}>0$
\end{tabular}
\\[7pt]

$f(\mathcal R,\mathcal Q)$
&
No additional poles
&
No pole condition
\\[4pt]

$f(R,\mathcal Q)$
&
At most one scalar pole
&
If present, $f^{(0)}_{,RR}>0$
\\[4pt]

$f(R,\hat Q,Q,\mathcal Q)$
&
At most one scalar pole
&
\begin{tabular}[t]{@{}l@{}}
$\left(f^{(0)}_{,\hat Q}\right)^2
-4f^{(0)}_{,\mathcal Q}f^{(0)}_{,Q}=0;$\\
if the scalar pole is present, $f^{(0)}_{,RR}>0$
\end{tabular}
\\[7pt]

$f(R,\mathcal R,\hat Q,Q,\mathcal Q)$,
$\sigma_2\neq0$
&
Two scalar poles
&
\begin{tabular}[t]{@{}l@{}}
$f^{(0)}_{,\hat Q}=-2f^{(0)}_{,Q},\quad
 f^{(0)}_{,\mathcal Q}=f^{(0)}_{,Q},$\\
$\hyperlink{78}{\lambda_\Psi>0},\quad
 \hyperlink{78}{\omega_\Psi>2\sqrt{\lambda_\Psi}},\quad
 f^{(0)}_{,\mathcal R}<0$
\end{tabular}
\\[7pt]

$f(R,\mathcal R,\hat Q,Q,\mathcal Q)$,
$\sigma_2=0$
&
At most one scalar and one spin--2 pole
&
\begin{tabular}[t]{@{}l@{}}
$\mathcal C^2-4\mathcal D\mathcal E=0;$\\
scalar pole:
$\hyperlink{sigmas}{\sigma_1<0},\quad
 \hyperlink{reduced-scalar-mass-residue}{\rho_0>0};$\\
spin--2 pole, if present:
$\hyperlink{53}{\omega_\Phi<0},\quad
 \hyperlink{spin2-single-residue}{\rho_2>0}$
\end{tabular}
\\

\bottomrule
\end{tabularx}

\vspace{3pt}
\begin{minipage}{\textwidth}
\footnotesize
\textit{Scope and domain of validity.}
The table summarizes finite poles in the metric propagator.
Affine modes are analyzed separately in
Appendix~\ref{app:affine-modes}. The general
$f(R,\mathcal R,\hat Q,Q,\mathcal Q)$ results assume
$f^{(0)}_{,\mathcal R}\neq0$ and
$3(\mathcal A+\mathcal B)+\mathcal C+2\mathcal D\neq0$.
For $f(R,\mathcal R)$, these conditions reduce to
$f^{(0)}_{,\mathcal R}\neq0$ and
$f^{(0)}_{,R\mathcal R}
+f^{(0)}_{,\mathcal R\mathcal R}\neq0$.
The $f(R,\hat Q,Q,\mathcal Q)$ and $f(R,\mathcal Q)$ rows refer to the
$f^{(0)}_{,\mathcal Q}\neq0$ branches with asymptotically flat
perturbations, as analyzed directly in
Appendix~\ref{APPENDIX-B}. These are domain assumptions rather than
additional stability conditions.
\end{minipage}
\end{table}
\FloatBarrier

\section{Conclusions}
\label{conclusions}

We have investigated extended hybrid metric--Palatini theories of
gravity defined by the action \eqref{actionf}. Their Lagrangian depends
on the metric and Palatini curvature scalars together with the
corresponding quadratic Ricci invariants. We derived the metric and
connection field equations and studied their weak--field limit around
Minkowski spacetime. For the class of theories considered here, the torsion tensor is entirely determined by its trace. Projective invariance allows this trace to be set to zero without affecting the dynamics. The independent connection perturbation is obtained through the auxiliary-metric construction on the branch {\small$\smash{f^{(0)}_{,\mathcal R}\neq0}$}. Since this construction is not applicable when {\small$f^{(0)}_{,\mathcal R}=0$}, the corresponding subclasses are analyzed in Appendix~\ref{APPENDIX-B}.

The linearized metric equations were cast in operator form and inverted
using the standard spin projectors. The resulting metric propagator in
the conserved-source sector determines the finite spin--2 and spin--0
poles that couple to the metric source. When the tensor denominator is
quadratic and its two roots remain as distinct poles, at least one
spin--2 residue is negative. Reducing the tensor denominator to at most
one additional pole leads to two distinct scalar branches.

In the five invariants case, for the case containing two real and distinct scalar poles,
the absence of ghostlike metric-coupled poles requires the spin--2
correction to vanish. On the
 {\small$\smash{f^{(0)}_{,\mathcal R}\neq0}$} branch, this occurs when
 {\small$\smash{f^{(0)}_{,\hat Q}=-2f^{(0)}_{,Q}}$} and
 {\small$\smash{f^{(0)}_{,\mathcal Q}=f^{(0)}_{,Q}}$}. The remaining additional poles in
the metric propagator are then scalar, and their stability is determined
by their squared masses and residues. On the reduced-order scalar
branch, the scalar denominator is linear in $k^2$ and contains at most
one additional finite pole. A positive-residue spin--2 pole may coexist
with this scalar, provided that the stability conditions (see Table~\ref{tab:pole-summary}) for both poles
are satisfied independently.

However, the metric propagator does not contain all the modes of the full metric--connection system. As shown in
Appendix~\ref{app:affine-modes}, the conditions that make the spin--2
correction vanish imply $\mathcal C=-2\mathcal D$. For
$\mathcal D\neq0$, the complete system therefore retains a purely
affine spin--2 mode with
$m_{\mathrm{aff},2}^{2}=1/\mathcal D$. This mode has zero residue in the metric propagator. It is
non-tachyonic when $\mathcal D>0$, but the coupled field equations alone do not determine whether the mode is healthy or ghostlike. If $\mathcal D=0$, no finite
affine spin--2 pole remains. The complementary surface
$3(\mathcal A+\mathcal B)+\mathcal C+2\mathcal D=0$ may similarly
support a purely affine scalar mode when a finite root exists. The
finite purely affine branches are summarized in
Table~\ref{tab:appC-affine-summary}.

For generalized hybrid $f(R,\mathcal R)$ gravity, the generic branch
contains two additional metric-coupled scalar poles. Their stability
conditions can be expressed directly in terms of the background
derivatives of $f$. The reduced-order condition
 {\small$\smash{\left(f^{(0)}_{,R\mathcal R}\right)^2
=f^{(0)}_{,\mathcal R\mathcal R}f^{(0)}_{,RR}}$} leaves at most one
finite metric-coupled scalar pole. By contrast, the complementary
surface
 {\small$\smash{f^{(0)}_{,R\mathcal R}+f^{(0)}_{,\mathcal R\mathcal R}=0}$} lies outside
the domain of the general propagator reduction and may support the
purely affine scalar mode discussed in
Appendix~\ref{app:affine-modes}.

Several relevant subclasses follow as consistent limits of the general
framework. Metric $f(R)$ and mixed $f(R,\mathcal Q)$ gravity contain at
most one additional scalar pole, which is non-tachyonic and has positive
residue when  {\small$\smash{f^{(0)}_{,RR}>0}$}. On the
 {\small$\smash{f^{(0)}_{,\mathcal Q}\neq0}$} branch of
$f(R,\hat Q,Q,\mathcal Q)$ gravity, the condition
 {\small$\smash{\left(f^{(0)}_{,\hat Q}\right)^2
-4f^{(0)}_{,\mathcal Q}f^{(0)}_{,Q}=0}$} removes the ghostlike metric-coupled spin--2 pole. The remaining scalar, when present, is
stable for  {\small$\smash{f^{(0)}_{,RR}>0}$}. On this branch, the connection equation
fixes the Palatini Ricci tensor algebraically in terms of the metric
Ricci tensor, and no finite purely affine symmetric-Ricci pole is
generated. Quadratic Palatini $f(\mathcal R,\mathcal Q)$ gravity and the Palatini
$f(\mathcal R)$ limit contain no additional finite curvature poles
around Minkowski spacetime. Table~\ref{tab:viability} summarizes the
additional metric-coupled poles and necessary stability conditions for
the general theory and its relevant subclasses.

The conditions derived from the metric propagator should therefore be
understood as necessary flat-space stability criteria for the
metric-coupled sector. When a purely affine mode is present, determining
its kinetic sign requires a separate analysis of the quadratic action
for the connection perturbations. Moreover, the present conditions do
not establish stability on curved backgrounds or beyond the linearized
regime. Extending the analysis to such regimes remains an important
direction for future work
\cite{Belenchia2018,BELTRANJIMENEZ20181}.

\section*{Acknowledgements}

The authors are grateful to Santiago Esteban Perez Bergliaffa for
valuable discussions, comments, and guidance throughout this work.
They also thank Gonzalo J. Olmo for helpful suggestions that improved
the manuscript and Jos\'e Beltr\'an Jim\'enez for useful correspondence
and insightful comments on possible nonlinear extensions of the
stability analysis.

\section*{Statements and Declarations}

\noindent\textbf{Funding.}
This study was financed in part by the Coordena\c{c}\~ao de
Aperfei\c{c}oamento de Pessoal de N\'ivel Superior -- Brasil
(CAPES) -- Finance Code 001.

\noindent\textbf{Competing interests.}
The authors have no relevant financial or non-financial interests
to disclose.

\noindent\textbf{Data availability.}
No datasets were generated or analysed during the current study.

\begin{appendices}

\section{Pure-trace nature of torsion in the symmetric-Ricci sector}
\label{APPENDIX-A}

In this Appendix, we reproduce the main steps of the derivation
presented in Ref.~\cite{Alfonso_2017} and adapt them to the extended
hybrid metric--Palatini framework considered here. The effect of the
Ricci--squared invariants is contained in the explicit form of the
tensor that multiplies the variation of the Palatini Ricci tensor.

In a metric--affine setting, the independent connection
$\widetilde{\Gamma}^{\lambda}{}_{\mu\nu}$ is not assumed to be
symmetric. Its torsion tensor is defined as
\begin{equation}
S^{\lambda}{}_{\mu\nu}
\equiv
\widetilde{\Gamma}^{\lambda}{}_{[\mu\nu]}.
\end{equation}
The corresponding Palatini Ricci tensor
$\mathcal{R}_{\mu\nu}(\widetilde{\Gamma})$ is therefore not symmetric
in general. In this work, however, the action depends on the
independent connection only through the symmetric part
$\mathcal{R}_{(\mu\nu)}$. The connection variation can then be written
as
\begin{equation}
\delta_{\widetilde{\Gamma}}S
=
\frac{1}{2\kappa^2}
\int d^4x\,\sqrt{-g}\,
q^{\mu\nu}\delta\mathcal{R}_{(\mu\nu)},
\end{equation}
where, for the Lagrangian
$f(R,\mathcal{R},\hat{Q},Q,\mathcal{Q})$,
\begin{equation}
q^{\mu\nu}
\equiv
\frac{\partial f}{\partial\mathcal{R}_{(\mu\nu)}}
=
f_{,\mathcal{R}}g^{\mu\nu}
+f_{,\hat{Q}}R^{\mu\nu}
+2f_{,\mathcal{Q}}\mathcal{R}^{(\mu\nu)}.
\end{equation}
All indices are raised with $g^{\mu\nu}$. Since the dependence on the
Palatini Ricci tensor is restricted to
$\mathcal{R}_{(\mu\nu)}$, the tensor $q^{\mu\nu}$ is symmetric:
\begin{equation}
q^{[\mu\nu]}=0.
\label{eq:Zsymmetric}
\end{equation}

In the presence of torsion, the variation of the Palatini Ricci tensor
is
\begin{equation}
\begin{aligned}
\delta\mathcal{R}_{\mu\nu}
={}&
\widetilde{\nabla}_{\lambda}
\left(
\delta\widetilde{\Gamma}^{\lambda}{}_{\nu\mu}
\right)
-\widetilde{\nabla}_{\nu}
\left(
\delta\widetilde{\Gamma}^{\lambda}{}_{\lambda\mu}
\right)
\\
&+
2S^{\lambda}{}_{\rho\nu}
\delta\widetilde{\Gamma}^{\rho}{}_{\lambda\mu},
\end{aligned}
\label{eq:deltaRicciTorsion}
\end{equation}
where $\widetilde{\nabla}$ denotes the covariant derivative associated
with $\widetilde{\Gamma}^{\lambda}{}_{\mu\nu}$. After integration by
parts, the connection field equations take the form
\begin{equation}
\begin{aligned}
&2\sqrt{-g}
\left(
q^{\mu\sigma}S^{\nu}{}_{\alpha\sigma}
+q^{\mu\nu}S^{\sigma}{}_{\sigma\alpha}
-\delta^{\nu}_{\alpha}
 q^{\mu\rho}S^{\sigma}{}_{\sigma\rho}
\right)
\\
&\quad
+\delta^{\nu}_{\alpha}
 \widetilde{\nabla}_{\sigma}
 \left(
 \sqrt{-g}\,q^{\mu\sigma}
 \right)
-\widetilde{\nabla}_{\alpha}
 \left(
 \sqrt{-g}\,q^{\mu\nu}
 \right)
=0,
\end{aligned}
\label{eq:connEOM_pretrace}
\end{equation}
where
\begin{equation}
S_\mu
\equiv
S^{\lambda}{}_{\lambda\mu}.
\end{equation}

Tracing Eq.~\eqref{eq:connEOM_pretrace} over $\nu$ and $\alpha$ gives
\begin{equation}
\widetilde{\nabla}_{\sigma}
\left(
\sqrt{-g}\,q^{\mu\sigma}
\right)
=
\frac{4}{3}\sqrt{-g}\,
S_{\sigma}q^{\mu\sigma}.
\label{eq:trace_connEOM}
\end{equation}
Substituting Eq.~\eqref{eq:trace_connEOM} back into
Eq.~\eqref{eq:connEOM_pretrace}, we obtain
\begin{equation}
\begin{aligned}
\widetilde{\nabla}_{\alpha}
\left(
\sqrt{-g}\,q^{\mu\nu}
\right)
=
2\sqrt{-g}
\left(
q^{\mu\sigma}S^{\nu}{}_{\alpha\sigma}
+q^{\mu\nu}S_{\alpha}
-\frac{1}{3}\delta^{\nu}_{\alpha}
 q^{\mu\rho}S_{\rho}
\right).
\end{aligned}
\label{eq:connEOM_posttrace}
\end{equation}

We decompose the independent connection as
\begin{equation}
\widetilde{\Gamma}^{\alpha}{}_{\mu\nu}
=
C^{\alpha}{}_{\mu\nu}
+
S^{\alpha}{}_{\mu\nu},
\qquad
C^{\alpha}{}_{\mu\nu}
\equiv
\widetilde{\Gamma}^{\alpha}{}_{(\mu\nu)}.
\end{equation}
Equation~\eqref{eq:connEOM_posttrace} then becomes
\begin{equation}
\begin{aligned}
\frac{1}{\sqrt{-g}}
\nabla^{C}_{\alpha}
\left(
\sqrt{-g}\,q^{\mu\nu}
\right)
={}&
q^{\mu\sigma}S^{\nu}{}_{\alpha\sigma}
+q^{\mu\nu}S_{\alpha}
-q^{\sigma\nu}S^{\mu}{}_{\alpha\sigma}
\\
&-
\frac{2}{3}\delta^{\nu}_{\alpha}
q^{\mu\rho}S_{\rho}.
\end{aligned}
\label{205}
\end{equation}

We now introduce the transformed connection
\begin{equation}
\widehat{\Gamma}^{\alpha}{}_{\mu\nu}
=
\widetilde{\Gamma}^{\alpha}{}_{\mu\nu}
+\frac{2}{3}\delta^{\alpha}_{\nu}S_{\mu},
\end{equation}
whose symmetric and antisymmetric parts are
\begin{equation}
\widehat{C}^{\lambda}{}_{\mu\nu}
=
C^{\lambda}{}_{\mu\nu}
+\frac{1}{3}
\left(
\delta^{\lambda}_{\nu}S_{\mu}
+\delta^{\lambda}_{\mu}S_{\nu}
\right),
\end{equation}
and
\begin{equation}
\widehat{S}^{\lambda}{}_{\mu\nu}
=
S^{\lambda}{}_{\mu\nu}
+\frac{1}{3}
\left(
\delta^{\lambda}_{\nu}S_{\mu}
-\delta^{\lambda}_{\mu}S_{\nu}
\right).
\end{equation}
By construction,
$\widehat{S}^{\lambda}{}_{\lambda\mu}=0$. In terms of these variables,
Eq.~\eqref{205} becomes
\begin{equation}
\begin{aligned}
\frac{1}{\sqrt{-g}}
\nabla^{\widehat{C}}_{\alpha}
\left(
\sqrt{-g}\,q^{\mu\nu}
\right)
=
q^{\mu\sigma}\widehat{S}^{\nu}{}_{\alpha\sigma}
-q^{\nu\sigma}\widehat{S}^{\mu}{}_{\alpha\sigma}.
\end{aligned}
\end{equation}

We now assume that $q^{\mu\nu}$ is nondegenerate. Let
$q_{\mu\nu}$ denote its inverse and define
$q\equiv\det(q_{\mu\nu})$. Following Ref.~\cite{Alfonso_2017}, the
previous equation can be written as
\begin{equation}
\left(
\widehat{\nabla}_{\alpha}
+V_{\alpha}
\right)q^{\mu\nu}
=
2q^{\mu\sigma}
\widehat{S}^{\nu}{}_{\alpha\sigma},
\label{A15}
\end{equation}
where
\begin{equation}
V_{\alpha}
\equiv
\partial_{\alpha}
\ln\left(
\frac{\sqrt{-q}}{\sqrt{-g}}
\right),
\end{equation}
and $\widehat{\nabla}_{\alpha}$ is the covariant derivative associated
with $\widehat{\Gamma}^{\alpha}{}_{\mu\nu}$.

Introducing an auxiliary metric $\tilde{g}_{\mu\nu}$ through
\begin{equation}
q^{\mu\nu}
=
\frac{\sqrt{-g}}{\sqrt{-q}}\,
\tilde{g}^{\mu\nu},
\end{equation}
removes $V_\alpha$ from Eq.~\eqref{A15} and gives
\begin{equation}
\widehat{\nabla}_{\alpha}\tilde{g}^{\mu\nu}
=
2\tilde{g}^{\mu\sigma}
\widehat{S}^{\nu}{}_{\alpha\sigma}.
\label{A18}
\end{equation}
After lowering the indices and combining the corresponding cyclic
permutations, we obtain
\begin{equation}
\begin{aligned}
\partial_{\alpha}\tilde{g}_{\mu\nu}
+\partial_{\nu}\tilde{g}_{\alpha\mu}
-\partial_{\mu}\tilde{g}_{\nu\alpha}
=
2\tilde{g}_{\mu\sigma}
\widehat{\Gamma}^{\sigma}{}_{\alpha\nu}.
\end{aligned}
\label{A19}
\end{equation}

The left-hand side of Eq.~\eqref{A19} is symmetric under the exchange
$\alpha\leftrightarrow\nu$. It follows that the antisymmetric part of
$\widehat{\Gamma}^{\sigma}{}_{\alpha\nu}$ vanishes:
\begin{equation}
\widehat{S}^{\sigma}{}_{\alpha\nu}=0.
\end{equation}
Its symmetric part is therefore the Levi--Civita connection of the
auxiliary metric,
\begin{equation}
\widehat{C}^{\sigma}{}_{\alpha\nu}
=
\frac{1}{2}\tilde{g}^{\mu\sigma}
\left(
\partial_{\alpha}\tilde{g}_{\mu\nu}
+\partial_{\nu}\tilde{g}_{\alpha\mu}
-\partial_{\mu}\tilde{g}_{\nu\alpha}
\right).
\end{equation}
The original connection can consequently be written as
\begin{equation}
\widetilde{\Gamma}^{\alpha}{}_{\mu\nu}
=
\widehat{C}^{\alpha}{}_{\mu\nu}
-\frac{2}{3}\delta^{\alpha}_{\nu}S_{\mu},
\end{equation}
and its torsion is determined entirely by its trace:
\begin{equation}
S^{\lambda}{}_{\mu\nu}
=
\frac{1}{3}
\left(
\delta^{\lambda}_{\mu}S_{\nu}
-\delta^{\lambda}_{\nu}S_{\mu}
\right).
\end{equation}

Thus, whenever $q^{\mu\nu}$ admits an inverse, the torsion tensor is
determined entirely by the undetermined trace $S_\mu$. This trace does
not contribute to $\mathcal{R}_{(\mu\nu)}$ and hence does not enter the
action or the metric field equations
\cite{Alfonso_2017,Olmo:2011}. It may therefore be fixed by setting
$S_\mu=0$, in which case the torsion tensor vanishes.

\section{Direct weak--field analysis of the
\texorpdfstring{$f(R,\hat Q,Q,\mathcal Q)$}{f(R,Qhat,Q,Qcal)}
and \texorpdfstring{$f(R,\mathcal Q)$}{f(R,Qcal)} subclasses}
\label{APPENDIX-B}

In the weak--field construction of Sec.~\ref{Weak-field-limit}, the
zeroth-order value of Eq.~\eqref{eq:q} on the Minkowski background is
\begin{equation}
q^{(0)\mu\nu}
=
f_{,\mathcal R}^{(0)}\eta^{\mu\nu}.
\end{equation}
That construction assumes $f_{,\mathcal R}^{(0)}\neq0$, so that
$q^{(0)\mu\nu}$ is nondegenerate and its inverse and determinant are
well defined. The $f(R,\hat Q,Q,\mathcal Q)$ and $f(R,\mathcal Q)$ subclasses considered in Secs.~\ref{A} and~\ref{sec:fRQcal}, respectively, satisfy $f_{,\mathcal R}=0$ and hence $q^{(0)\mu\nu}=0$. Consequently, the auxiliary-metric expansion developed in Sec.~\ref{Weak-field-limit} is not applicable to this branch. Their metric propagators and stability conditions are instead obtained from a direct linearization of the metric and connection field equations. This Appendix presents the complete derivation of the results reported in those sections.

\subsection{The \texorpdfstring{$f(R,\hat Q,Q,\mathcal Q)$}
{f(R,Qhat,Q,Qcal)} subclass}

For a function $f(R,\hat Q,Q,\mathcal Q)$, Eq.~\eqref{eq:q} reduces to
\begin{equation}
q^{\mu\nu}
=
f_{,\hat Q}R^{\mu\nu}
+
2f_{,\mathcal Q}\mathcal R^{(\mu\nu)}.
\label{eq:appB-q}
\end{equation}
Its first-order perturbation around Minkowski spacetime is therefore
\begin{equation}
q^{(1)\mu\nu}
=
f_{,\hat Q}^{(0)}R^{(1)\mu\nu}
+
2f_{,\mathcal Q}^{(0)}
\mathcal R^{(1)(\mu\nu)}.
\label{eq:appB-q1}
\end{equation}
Corrections to $f_{,\hat Q}$ and $f_{,\mathcal Q}$ multiply curvature
tensors that are already of first perturbative order and hence do not
contribute at linear order.

We now linearize the general connection equation
\eqref{eq:83}. Since $q^{\mu\nu}$ has no zeroth-order contribution, the
terms proportional to products of $q^{\mu\nu}$ and the torsion tensor
are of second perturbative order. Moreover,
$\sqrt{-g}\,q^{\mu\nu}=q^{(1)\mu\nu}$ at linear order. The connection
equation therefore reduces to
\begin{equation}
\partial_{\alpha}
\left[
f_{,\hat Q}^{(0)}R^{(1)\mu\nu}
+
2f_{,\mathcal Q}^{(0)}
\mathcal R^{(1)(\mu\nu)}
\right]
=0.
\label{eq:appB-linear-connection}
\end{equation}

Equation~\eqref{eq:appB-linear-connection} implies
\begin{equation}
f_{,\hat Q}^{(0)}R^{(1)\mu\nu}
+
2f_{,\mathcal Q}^{(0)}
\mathcal R^{(1)(\mu\nu)}
=
\mathcal{Y}^{\mu\nu},
\label{eq:appB-constant-tensor}
\end{equation}
where $\mathcal{Y}^{\mu\nu}$ is spacetime independent. For asymptotically flat
perturbations, both linearized Ricci tensors vanish at spatial infinity,
and therefore $\mathcal{Y}^{\mu\nu}=0$. Hence
\begin{equation}
f_{,\hat Q}^{(0)}R^{(1)}_{\mu\nu}
+
2f_{,\mathcal Q}^{(0)}
\mathcal R^{(1)}_{(\mu\nu)}
=0.
\label{eq:appB-Ricci-relation}
\end{equation}
For $f_{,\mathcal Q}^{(0)}\neq0$, this gives
\begin{equation}
\mathcal R^{(1)}_{(\mu\nu)}
=
-\frac{f_{,\hat Q}^{(0)}}
{2f_{,\mathcal Q}^{(0)}}
R^{(1)}_{\mu\nu}.
\label{eq:appB-Ricci-solution}
\end{equation}
Equation~\eqref{eq:appB-Ricci-solution} allows
$\mathcal R^{(1)}_{(\mu\nu)}$ to be eliminated directly from the
linearized metric field equations.

Linearizing Eq.~\eqref{field-equation-Q} for the present subclass,
using $f_{,\mathcal R}=0$ and $f^{(0)}=0$, gives
\begin{align}
f_{,R}^{(0)}G^{(1)}_{\mu\nu}
&+
f_{,RR}^{(0)}
\left(
\eta_{\mu\nu}\Box
-\partial_{\mu}\partial_{\nu}
\right)R^{(1)}+
f_{,Q}^{(0)}
\left[
\Box R^{(1)}_{\mu\nu}
+\eta_{\mu\nu}
\partial_{\alpha}\partial_{\beta}
R^{(1)\alpha\beta}
-2\partial^{\lambda}\partial_{(\mu}
R^{(1)}_{\nu)\lambda}
\right]
\nonumber\\
&+
\frac{f_{,\hat Q}^{(0)}}{2}
\left[
\Box\mathcal R^{(1)}_{(\mu\nu)}
+\eta_{\mu\nu}
\partial_{\alpha}\partial_{\beta}
\mathcal R^{(1)(\alpha\beta)}
-\partial^{\lambda}\partial_{\mu}
\mathcal R^{(1)}_{(\lambda\nu)}
-\partial^{\lambda}\partial_{\nu}
\mathcal R^{(1)}_{(\lambda\mu)}
\right]
=
\kappa^2T^{(1)}_{\mu\nu}.
\label{eq:appB-linear-metric}
\end{align}

Substituting Eq.~\eqref{eq:appB-Ricci-solution} and using the
linearized contracted Bianchi identity,
$\partial^\mu R^{(1)}_{\mu\nu}
=\frac{1}{2}\partial_\nu R^{(1)}$, we obtain
\begin{align}
f_{,R}^{(0)}G^{(1)}_{\mu\nu}
&+
f_{,RR}^{(0)}
\left(
\eta_{\mu\nu}\Box
-\partial_{\mu}\partial_{\nu}
\right)R^{(1)}
\nonumber\\
&+
\left[
f_{,Q}^{(0)}
-\frac{\left(f_{,\hat Q}^{(0)}\right)^2}
{4f_{,\mathcal Q}^{(0)}}
\right]
\left[
\Box R^{(1)}_{\mu\nu}
+\frac{1}{2}\eta_{\mu\nu}\Box R^{(1)}
-\partial_{\mu}\partial_{\nu}R^{(1)}
\right]
=
\kappa^2T^{(1)}_{\mu\nu}.
\label{eq:appB-closed-metric}
\end{align}

Imposing the normalization $f_{,R}^{(0)}=1$ adopted in
Sec.~\ref{A}, and inverting Eq.~\eqref{eq:appB-closed-metric} with the
same spin projectors used in Sec.~\ref{Constraints}, the functions
appearing in the propagator \eqref{propagator} are
\begin{equation}
\Phi(-k^2)
=
\frac{
f_{,Q}^{(0)}
-\dfrac{\left(f_{,\hat Q}^{(0)}\right)^2}
{4f_{,\mathcal Q}^{(0)}}
}{
1-
\left[
f_{,Q}^{(0)}
-\dfrac{\left(f_{,\hat Q}^{(0)}\right)^2}
{4f_{,\mathcal Q}^{(0)}}
\right]k^2
},
\label{eq:appB-Phi}
\end{equation}
and
\begin{equation}
\Psi(-k^2)
=
\frac{
3f_{,RR}^{(0)}
+2f_{,Q}^{(0)}
-\dfrac{\left(f_{,\hat Q}^{(0)}\right)^2}
{2f_{,\mathcal Q}^{(0)}}
}{
2\left\{
1+
\left[
3f_{,RR}^{(0)}
+2f_{,Q}^{(0)}
-\dfrac{\left(f_{,\hat Q}^{(0)}\right)^2}
{2f_{,\mathcal Q}^{(0)}}
\right]k^2
\right\}
}.
\label{eq:appB-Psi}
\end{equation}

If
\begin{equation}
f_{,Q}^{(0)}
-\frac{\left(f_{,\hat Q}^{(0)}\right)^2}
{4f_{,\mathcal Q}^{(0)}}
\neq0,
\end{equation}
the denominator of Eq.~\eqref{eq:appB-Phi} contains an additional
massive spin--2 pole. As in metric Ricci--squared gravity, its residue
has the opposite sign to that of the massless graviton and therefore
corresponds to a ghostlike excitation. Eliminating this pole requires
\begin{equation}
\left(f_{,\hat Q}^{(0)}\right)^2
-
4f_{,\mathcal Q}^{(0)}f_{,Q}^{(0)}
=0.
\label{eq:appB-ghost-free-condition}
\end{equation}
This reproduces Eq.~\eqref{condition-f(R,Q,Q,Q)}.

Under this condition,
Eqs.~\eqref{eq:appB-Phi} and \eqref{eq:appB-Psi} reduce to
\begin{equation}
\Phi(-k^2)=0,
\qquad
\Psi(-k^2)
=
\frac{3f_{,RR}^{(0)}}
{2\left(1+3f_{,RR}^{(0)}k^2\right)}.
\label{eq:appB-final-propagator}
\end{equation}
For $f_{,RR}^{(0)}\neq0$, the metric propagator contains an additional
scalar pole with mass
\begin{equation}
m^2
=
\left(3f_{,RR}^{(0)}\right)^{-1},
\end{equation}
and positive residue. It is non-tachyonic for
$f_{,RR}^{(0)}>0$. If $f_{,RR}^{(0)}=0$, the scalar correction
vanishes and no additional scalar pole is present.

\subsection{The \texorpdfstring{$f(R,\mathcal Q)$}{f(R,Qcal)} subclass}

The $f(R,\mathcal Q)$ subclass follows by setting
\begin{equation}
f_{,Q}^{(0)}
=
f_{,\hat Q}^{(0)}
=0.
\end{equation}
For $f_{,\mathcal Q}^{(0)}\neq0$,
Eq.~\eqref{eq:appB-linear-connection} becomes
\begin{equation}
\partial_{\alpha}
\mathcal R^{(1)(\mu\nu)}
=0.
\end{equation}
Asymptotic flatness therefore implies
\begin{equation}
\mathcal R^{(1)}_{(\mu\nu)}=0.
\end{equation}
The linearized metric equation reduces to
\begin{equation}
f_{,R}^{(0)}G^{(1)}_{\mu\nu}
+
f_{,RR}^{(0)}
\left(
\eta_{\mu\nu}\Box
-\partial_{\mu}\partial_{\nu}
\right)R^{(1)}
=
\kappa^2T^{(1)}_{\mu\nu},
\label{eq:appB-fRQcal-metric}
\end{equation}
which is the linearized metric $f(R)$ equation.

After imposing $f_{,R}^{(0)}=1$, the functions appearing in the
propagator \eqref{propagator} are given by
Eq.~\eqref{eq:appB-final-propagator}. For
$f_{,RR}^{(0)}\neq0$, the additional scalar pole has mass
\begin{equation}
m^2
=
\left(3f_{,RR}^{(0)}\right)^{-1},
\end{equation}
and positive residue. It is non-tachyonic for
$f_{,RR}^{(0)}>0$. If $f_{,RR}^{(0)}=0$, the scalar correction
vanishes.

The results obtained in this Appendix characterize the finite poles of
the metric propagator at the linearized level around Minkowski
spacetime. Affine modes are considered separately in
Appendix~\ref{app:affine-modes}. Assessing stability beyond the
flat-space linearized regime requires extending the analysis to curved
backgrounds and to the nonlinear dynamics.

\section{Affine modes of the linearized system}
\label{app:affine-modes}

The propagator derived in Sec.~\ref{Constraints} describes the response of the metric perturbation $h_{\mu\nu}$ to the matter source $T^{(1)}_{\mu\nu}$ for the $f(R,\mathcal{R},\hat{Q},Q,\mathcal{Q})$ theory. Since matter is minimally coupled to $g_{\mu\nu}$, a mode carried entirely by the independent connection can be absent from the metric response while remaining an eigenmode of the complete coupled system. We identify these modes directly from the linearized metric and connection equations.

We first consider the branch $f^{(0)}_{,\mathcal R}\neq0$, for which
the coefficients
$\mathcal A,\mathcal B,\mathcal C,\mathcal D,\mathcal E$ introduced in
Sec.~\ref{Weak-field-limit} are well defined. The branch
$f^{(0)}_{,\mathcal R}=0$ is discussed separately below.

\subsection{Scalar sector}

Taking the trace of Eq.~\eqref{eq:Ricci-cal} gives
\begin{equation}
\left[
1+(3\mathcal B+2\mathcal D)\Box
\right]\mathcal R^{(1)}
=
\left[
1-(3\mathcal A+\mathcal C)\Box
\right]R^{(1)}.
\label{eq:appC-first-trace}
\end{equation}
The trace of the metric equation~\eqref{Expand-field-eqQ2}, with
$f^{(0)}=0$, is
\begin{equation}
\begin{aligned}
&
\left[
f^{(0)}_{,R}
-\left(
3f^{(0)}_{,RR}
+2f^{(0)}_{,\mathcal R}\mathcal E
\right)\Box
\right]R^{(1)}
\\
&\qquad
+
f^{(0)}_{,\mathcal R}
\left[
1-(3\mathcal A+\mathcal C)\Box
\right]\mathcal R^{(1)}
=
-\kappa^2T^{(1)}.
\end{aligned}
\label{eq:appC-second-trace}
\end{equation}

For finite $k^2\neq0$, the scalar part of the metric perturbation
satisfies
\begin{equation}
\mathcal P_s^0h_{\mu\nu}
=
\frac{1}{3k^2}
\left(
\eta_{\mu\nu}
-\frac{k_\mu k_\nu}{k^2}
\right)R^{(1)}.
\label{eq:appC-metric-scalar-amplitude}
\end{equation}
A purely affine scalar mode is therefore characterized by
\begin{equation}
R^{(1)}=0,
\qquad
\mathcal R^{(1)}\neq0.
\label{eq:appC-affine-scalar-vector}
\end{equation}
The homogeneous form of Eqs.~\eqref{eq:appC-first-trace} and
\eqref{eq:appC-second-trace} then requires
\begin{equation}
\left[
1+(3\mathcal B+2\mathcal D)\Box
\right]\mathcal R^{(1)}=0,
\qquad
\left[
1-(3\mathcal A+\mathcal C)\Box
\right]\mathcal R^{(1)}=0.
\end{equation}
These equations admit a common finite root when
\begin{equation}
3(\mathcal A+\mathcal B)
+\mathcal C+2\mathcal D=0.
\label{eq:appC-affine-scalar-condition}
\end{equation}
This is the complementary surface to the second condition in
Eq.~\eqref{general-branch-assumptions}. Provided
$3\mathcal B+2\mathcal D\neq0$, the corresponding squared mass is
\begin{equation}
m_{\mathrm{aff},0}^{2}
=
-\frac{1}{3\mathcal B+2\mathcal D}
=
\frac{1}{3\mathcal A+\mathcal C}.
\label{eq:appC-affine-scalar-mass}
\end{equation}
The mode is non-tachyonic when
$3\mathcal B+2\mathcal D<0$. If
$3\mathcal B+2\mathcal D=0$, condition
\eqref{eq:appC-affine-scalar-condition} also gives
$3\mathcal A+\mathcal C=0$, and no finite affine scalar root remains.

On the affine surface~\eqref{eq:appC-affine-scalar-condition},
Eq.~\eqref{eq:appC-first-trace} becomes
\begin{equation}
\left[
1+(3\mathcal B+2\mathcal D)\Box
\right]
\left(
\mathcal R^{(1)}-R^{(1)}
\right)
=0.
\label{eq:appC-affine-scalar-difference}
\end{equation}
Away from the affine root, this equation gives
$\mathcal R^{(1)}=R^{(1)}$. At the root, it imposes no relation between
the two curvatures and admits the purely affine solution
\eqref{eq:appC-affine-scalar-vector}.

Using Eq.~\eqref{eq:appC-affine-scalar-difference} in the metric trace
equation, together with the normalization
$f^{(0)}_{,R}+f^{(0)}_{,\mathcal R}=1$, gives
\begin{equation}
\Bigg\{1+\Big[
f^{(0)}_{,\mathcal R}
(3\mathcal B+2\mathcal D)
-3f^{(0)}_{,RR}
-2f^{(0)}_{,\mathcal R}\mathcal E
\Big]\Box
\Bigg\}R^{(1)}
=
-\kappa^2T^{(1)}.
\label{eq:appC-affine-scalar-metric-response}
\end{equation}
The affine factor is absent from the sourced metric equation, although
the complete scalar operator factorizes as
\begin{align}
-\frac{1}{2}
\left[
a(\Box)-3c(\Box)
\right]
={}&
\left[
1+(3\mathcal B+2\mathcal D)\Box
\right]
\nonumber\\
&\times
\Bigg\{
1+
\Big[
f^{(0)}_{,\mathcal R}
(3\mathcal B+2\mathcal D)
-3f^{(0)}_{,RR}
-2f^{(0)}_{,\mathcal R}\mathcal E
\Big]\Box
\Bigg\}.
\label{eq:appC-affine-scalar-factorization}
\end{align}
Thus, the affine eigenvalue remains a zero of the complete coupled
system.

The scalar part of the metric propagator is
\begin{equation}
\Pi_0(k)
=
-\frac{\mathcal P_s^0}
{
2k^2
\left\{
1+
\left[
3f^{(0)}_{,RR}
+2f^{(0)}_{,\mathcal R}\mathcal E
-f^{(0)}_{,\mathcal R}
 (3\mathcal B+2\mathcal D)
\right]k^2
\right\}
}.
\label{eq:appC-scalar-metric-propagator-reduced}
\end{equation}
For noncoincident roots, this expression is finite at
$k^2=-m_{\mathrm{aff},0}^2$. The purely affine scalar therefore has
zero metric-source residue. This expresses its decoupling from the
metric source but does not determine the sign of its affine kinetic
term.

If
\begin{equation}
3f^{(0)}_{,RR}
+2f^{(0)}_{,\mathcal R}\mathcal E
-f^{(0)}_{,\mathcal R}
 (3\mathcal B+2\mathcal D)
\neq0,
\end{equation}
the remaining metric-coupled scalar has
\begin{equation}
m_{\mathrm{met},0}^{2}
=
\frac{1}{
3f^{(0)}_{,RR}
+2f^{(0)}_{,\mathcal R}\mathcal E
-f^{(0)}_{,\mathcal R}
 (3\mathcal B+2\mathcal D)
}.
\label{eq:appC-metric-scalar-mass}
\end{equation}
For a simple pole, its residue coefficient is $+\frac12$, and it is
non-tachyonic when the denominator in
Eq.~\eqref{eq:appC-metric-scalar-mass} is positive.

The affine and metric-coupled scalar roots coincide when
\begin{equation}
f^{(0)}_{,R}
(3\mathcal B+2\mathcal D)
+3f^{(0)}_{,RR}
+2f^{(0)}_{,\mathcal R}\mathcal E
=0.
\label{eq:appC-affine-scalar-coincident}
\end{equation}
On this degenerate surface, the separate attribution of residues to a
purely affine mode and a metric-coupled mode is no longer applicable.

%------------------------------------------------------------
\subsection{Spin--\texorpdfstring{$2$}{2} sector}
%------------------------------------------------------------

After separating the scalar traces, the spin--$2$ parts of the
connection and metric equations are
\begin{equation}
(1-\mathcal D\Box)
\mathcal P^2\mathcal R^{(1)}_{(\mu\nu)}
=
\left(
1+\frac{\mathcal C}{2}\Box
\right)
\mathcal P^2R^{(1)}_{\mu\nu},
\label{eq:appC-first-spin2}
\end{equation}
and
\begin{equation}
\left(
f^{(0)}_{,R}
+f^{(0)}_{,\mathcal R}\mathcal E\Box
\right)
\mathcal P^2R^{(1)}_{\mu\nu}
+
f^{(0)}_{,\mathcal R}
\left(
1+\frac{\mathcal C}{2}\Box
\right)
\mathcal P^2\mathcal R^{(1)}_{(\mu\nu)}
=
\kappa^2\mathcal P^2T^{(1)}_{\mu\nu}.
\label{eq:appC-second-spin2}
\end{equation}

For finite momentum,
\begin{equation}
\mathcal P^2R^{(1)}_{\mu\nu}
=
-\frac{1}{2}\Box\mathcal P^2h_{\mu\nu}.
\label{eq:appC-Ricci-h-spin2}
\end{equation}
A purely affine spin--$2$ mode therefore satisfies
\begin{equation}
\mathcal P^2R^{(1)}_{\mu\nu}=0,
\qquad
\mathcal P^2\mathcal R^{(1)}_{(\mu\nu)}\neq0.
\label{eq:appC-affine-spin2-vector}
\end{equation}
The homogeneous equations become
\begin{equation}
(1-\mathcal D\Box)
\mathcal P^2\mathcal R^{(1)}_{(\mu\nu)}=0,
\qquad
\left(
1+\frac{\mathcal C}{2}\Box
\right)
\mathcal P^2\mathcal R^{(1)}_{(\mu\nu)}=0.
\end{equation}
They admit a common finite root when
\begin{equation}
\mathcal C=-2\mathcal D.
\label{eq:appC-affine-spin2-condition}
\end{equation}
For $\mathcal D\neq0$, its squared mass is
\begin{equation}
m_{\mathrm{aff},2}^{2}
=
\frac{1}{\mathcal D}
=
-\frac{2}{\mathcal C},
\label{eq:appC-affine-spin2-mass}
\end{equation}
and the mode is non-tachyonic when $\mathcal D>0$. If
$\mathcal D=0$, Eq.~\eqref{eq:appC-affine-spin2-condition} also gives
$\mathcal C=0$, and no finite affine spin--$2$ root remains.

On the affine branch~\eqref{eq:appC-affine-spin2-condition},
Eq.~\eqref{eq:appC-first-spin2} becomes
\begin{equation}
(1-\mathcal D\Box)
\left(
\mathcal P^2\mathcal R^{(1)}_{(\mu\nu)}
-\mathcal P^2R^{(1)}_{\mu\nu}
\right)
=0.
\label{eq:appC-affine-spin2-difference}
\end{equation}
Away from the affine root, the projected metric and Palatini Ricci
tensors coincide. At the root, Eq.~\eqref{eq:appC-affine-spin2-difference}
admits the purely affine solution
\eqref{eq:appC-affine-spin2-vector}.

Substituting Eq.~\eqref{eq:appC-affine-spin2-difference} into the
projected metric equation gives
\begin{equation}
\left[
1+
f^{(0)}_{,\mathcal R}
(\mathcal E-\mathcal D)\Box
\right]
\mathcal P^2R^{(1)}_{\mu\nu}
=
\kappa^2\mathcal P^2T^{(1)}_{\mu\nu}.
\label{eq:appC-affine-spin2-metric-response}
\end{equation}
The affine factor again cancels from the metric response, whereas the
complete spin--$2$ operator factorizes as
\begin{equation}
a(\Box)
=
(1-\mathcal D\Box)
\left[
1+
f^{(0)}_{,\mathcal R}
(\mathcal E-\mathcal D)\Box
\right].
\label{eq:appC-affine-spin2-factorization}
\end{equation}

The spin--$2$ part of the metric propagator is consequently
\begin{equation}
\Pi_2(k)
=
\frac{\mathcal P^2}
{
k^2
\left[
1+
f^{(0)}_{,\mathcal R}
(\mathcal D-\mathcal E)k^2
\right]
}.
\label{eq:appC-spin2-metric-propagator-reduced}
\end{equation}
For noncoincident roots, $\Pi_2(k)$ is finite at
$k^2=-m_{\mathrm{aff},2}^2$. The purely affine spin--$2$ mode therefore
has zero metric-source residue, but its affine kinetic sign cannot be
inferred from the metric propagator.

For $\mathcal E\neq\mathcal D$, the remaining metric-coupled spin--$2$
pole has
\begin{equation}
m_{\mathrm{met},2}^{2}
=
\frac{1}{
f^{(0)}_{,\mathcal R}
(\mathcal D-\mathcal E)
}.
\label{eq:appC-metric-spin2-mass}
\end{equation}
For a simple pole, its residue coefficient is $-1$, so this mode is
ghostlike. It is absent when $\mathcal E=\mathcal D$.

Using the definitions below
Eq.~\eqref{Phi-particulaQ}, one finds
\begin{equation}
\delta_\Phi^2
-\omega_\Phi\delta_\Phi
+\lambda_\Phi
=
\frac{f^{(0)}_{,\mathcal R}}{4}
(\mathcal C+2\mathcal D)^2.
\label{eq:appC-Phi-cancellation}
\end{equation}
Since $f^{(0)}_{,\mathcal R}\neq0$, the common-factor condition is
equivalent to $\mathcal C=-2\mathcal D$.

The affine and metric-coupled spin--$2$ roots coincide when
\begin{equation}
\mathcal D
\left(
1-f^{(0)}_{,\mathcal R}
\right)
+f^{(0)}_{,\mathcal R}\mathcal E
=0.
\label{eq:appC-affine-spin2-coincident}
\end{equation}
On this degenerate surface, the separate attribution of residues to
the two spin--$2$ eigenmodes is no longer applicable.

Finally, the metric-propagator conditions
\eqref{spin2-ghost-free-conditions} imply
\begin{equation}
\mathcal C=-2\mathcal E,
\qquad
\mathcal D=\mathcal E,
\qquad
\mathcal C=-2\mathcal D.
\label{eq:appC-general-ghostfree-affine}
\end{equation}
Thus, $\mathcal E=\mathcal D$ removes the remaining metric-coupled
spin--$2$ pole, while $\mathcal C=-2\mathcal D$ places the complete
system on the affine spin--$2$ branch. For $\mathcal D\neq0$, these
conditions therefore leave a purely affine spin--$2$ mode with
$m_{\mathrm{aff},2}^{2}=1/\mathcal D$, which is non-tachyonic for
$\mathcal D>0$. If $\mathcal D=0$, one instead has
$\mathcal C=\mathcal E=0$, and no finite affine spin--$2$ root remains.

%------------------------------------------------------------
\subsection{Relevant subclasses and summary}
%------------------------------------------------------------

For the $f(R,\mathcal{R},\hat{Q},Q,\mathcal{Q})$ theory, on the branch $f^{(0)}_{,\mathcal R}\neq0$, the affine spin--$2$
condition and squared mass can be written as
\begin{equation}
f^{(0)}_{,\hat Q}
=
-2f^{(0)}_{,\mathcal Q},
\qquad
m_{\mathrm{aff},2}^{2}
=
\frac{f^{(0)}_{,\mathcal R}}
{f^{(0)}_{,\mathcal Q}},
\end{equation}
provided $f^{(0)}_{,\mathcal Q}\neq0$. The mode is non-tachyonic when
$f^{(0)}_{,\mathcal R}/f^{(0)}_{,\mathcal Q}>0$.

Similarly, the affine scalar condition is
\begin{equation}
3\left(
f^{(0)}_{,R\mathcal R}
+f^{(0)}_{,\mathcal R\mathcal R}
\right)
+f^{(0)}_{,\hat Q}
+2f^{(0)}_{,\mathcal Q}
=0,
\end{equation}
and its squared mass is
\begin{equation}
m_{\mathrm{aff},0}^{2}
=
-\frac{f^{(0)}_{,\mathcal R}}
{3f^{(0)}_{,\mathcal R\mathcal R}
+2f^{(0)}_{,\mathcal Q}},
\end{equation}
provided
$3f^{(0)}_{,\mathcal R\mathcal R}
+2f^{(0)}_{,\mathcal Q}\neq0$.

For generalized hybrid $f(R,\mathcal R)$ models,
$\mathcal C=\mathcal D=\mathcal E=0$, and hence there is no finite
affine spin--$2$ root. A purely affine scalar appears when
\begin{equation}
f^{(0)}_{,R\mathcal R}
+f^{(0)}_{,\mathcal R\mathcal R}
=0,
\label{eq:appC-hybrid-affine-scalar-condition}
\end{equation}
with
\begin{equation}
m_{\mathrm{aff},0}^{2}
=
-\frac{f^{(0)}_{,\mathcal R}}
{3f^{(0)}_{,\mathcal R\mathcal R}},
\label{eq:appC-hybrid-affine-scalar-mass}
\end{equation}
provided $f^{(0)}_{,\mathcal R\mathcal R}\neq0$. On the same surface,
the remaining metric-coupled scalar has
\begin{equation}
m_{\mathrm{met},0}^{2}
=
\frac{1}{
3\left(
f^{(0)}_{,RR}
-f^{(0)}_{,\mathcal R\mathcal R}
\right)
},
\end{equation}
with residue coefficient $+\frac12$. It is non-tachyonic when
$f^{(0)}_{,RR}>f^{(0)}_{,\mathcal R\mathcal R}$.

For the restricted $R+f(\mathcal R)$ model, $\mathcal A=0$, so the
affine condition requires $\mathcal B=0$, and no finite affine scalar
root remains. For quadratic Palatini $f(\mathcal R,\mathcal Q)$ models,
the scalar and spin--$2$ operators contain no finite curvature roots.
The same holds in the Palatini $f(\mathcal R)$ limit.

When $f^{(0)}_{,\mathcal R}=0$, the preceding formulas do not apply.
On the $f^{(0)}_{,\mathcal Q}\neq0$ branch of
$f(R,\hat Q,Q,\mathcal Q)$, Appendix~\ref{APPENDIX-B} gives
\begin{equation}
\mathcal R^{(1)}_{(\mu\nu)}
=
-\frac{f^{(0)}_{,\hat Q}}
{2f^{(0)}_{,\mathcal Q}}
R^{(1)}_{\mu\nu}
\label{eq:appC-fRQQQ-relation}
\end{equation}
for nonzero momentum, with the constant mode removed by asymptotic
flatness. The affine curvature is therefore fixed by the metric
curvature, and no finite purely affine symmetric-Ricci mode is
supported. Consequently, the condition
\begin{equation}
\left(
f^{(0)}_{,\hat Q}
\right)^2
-
4f^{(0)}_{,\mathcal Q}f^{(0)}_{,Q}
=0
\end{equation}
removes the massive spin--$2$ curvature pole on this branch rather
than converting it into a purely affine mode.

For the $f(R,\mathcal Q)$ subclass,
Eq.~\eqref{eq:appC-fRQQQ-relation} reduces to
$\mathcal R^{(1)}_{(\mu\nu)}=0$. The pure metric $f(R)$ limit has no
independent connection sector. If
$f^{(0)}_{,\mathcal R}=f^{(0)}_{,\mathcal Q}=0$ while the action
retains dependence on the independent connection, the linearized
connection equation is more degenerate and must be analyzed
separately.

The finite purely affine modes on the
$f^{(0)}_{,\mathcal R}\neq0$ branch are summarized in
Table~\ref{tab:appC-affine-summary}.

\begin{table}[!htbp]
\centering
\caption{Purely affine finite-curvature modes on the
$f^{(0)}_{,\mathcal R}\neq0$ branch.\\}
\label{tab:appC-affine-summary}
\setlength{\tabcolsep}{3pt}
\renewcommand{\arraystretch}{1.5}
\small
\rowcolors{2}{gray!10}{white}
\begin{tabularx}{\textwidth}{@{}
>{\raggedright\arraybackslash}p{0.2\textwidth}
>{\raggedright\arraybackslash}p{0.4\textwidth}
X@{}}
\toprule
\textbf{Mode}
& \textbf{Existence conditions and masses}
& \textbf{Non-tachyonic conditions} \\
\midrule

Affine spin--$2$
&
\begin{tabular}[t]{@{}l@{}}
$\mathcal C=-2\mathcal D,\quad\mathcal D\neq0;$\\
$m_{\mathrm{aff},2}^{2}=1/\mathcal D$
\end{tabular}
&
$\mathcal D>0$
\\[7pt]

Affine scalar
&
\begin{tabular}[t]{@{}l@{}}
$3(\mathcal A+\mathcal B)+\mathcal C+2\mathcal D=0,$\\
$3\mathcal B+2\mathcal D\neq0;$\\
$m_{\mathrm{aff},0}^{2}
=-1/(3\mathcal B+2\mathcal D)$
\end{tabular}
&
$3\mathcal B+2\mathcal D<0$
\\

\bottomrule
\end{tabularx}

\vspace{3pt}
\begin{minipage}{\textwidth}
\footnotesize
\textit{Scope.}
The table assumes finite, noncoincident roots on the branch
$f^{(0)}_{,\mathcal R}\neq0$. Both modes have zero metric-source
residue. Their affine kinetic signs are not determined by the metric
propagator, so the non-tachyonic conditions shown above do not establish
that the modes are ghost-free. For $f(R,\mathcal R)$ gravity, $\mathcal C=\mathcal D=\mathcal E=0$; therefore, no purely affine spin--$2$ mode exists, whereas a finite purely affine scalar mode may still be present.
\end{minipage}
\end{table}
\FloatBarrier

\end{appendices}

\bibliographystyle{unsrtnat}
\bibliography{bibliografia}

\end{document}